\documentclass[10pt, journal, paper, onecolumn]{IEEEtran}
\usepackage[noadjust]{cite}      
\usepackage{graphicx}  
\usepackage{amsmath,amssymb,epsfig}
\usepackage{multicol}
\usepackage{algorithm}
\usepackage{url}
\usepackage{float}
\usepackage{amsfonts}
\usepackage{theorem}

\makeatletter
    \newcommand\figcaption{\def\@captype{figure}\caption}
    \newcommand\tabcaption{\def\@captype{table}\caption}
\makeatother

\newtheorem{thm}{Theorem}
\newtheorem{cor}{Corollary}
\newtheorem{lem}{Lemma}

\theoremstyle{definition}
\newtheorem{defn}{Definition}
\theoremstyle{remark}
\newtheorem{rem}{Remark}

\begin{document}
\title{Algebraic Soft-Decision Decoding of Reed-Solomon Codes Using Bit-level Soft Information
\thanks{This work was supported in part by Seagate Research Inc., Pittsburgh, PA, USA and in part by
the Information Storage Industry Consortium. The material in this paper was presented in part at IEEE
International Symposium on Information Theory, Adelaide, Australia, 2005 and in part presented in $44^{th}$
annual Allerton conference on communication, control and computing, Monticello, Illinois 2006. Jing Jiang ($jingj@qualcomm.com$) was with Texas A\&M University and is now with Qualcomm Inc. Krishna~R.~Narayanan ($krn@ece.tamu.edu$) is with Texas A\&M University.}}
\author{Jing~Jiang and Krishna~R.~Narayanan\\
Department of Electrical and Computer Engineering, \\
Texas A\&M University,\\
College Station, TX, 77843, U.S.A} \maketitle{}

\begin{abstract}
The performance of algebraic soft-decision decoding of Reed-Solomon codes using bit-level soft information is
investigated. Optimal multiplicity assignment strategies of algebraic soft-decision decoding with infinite cost
are first studied over erasure channels and the binary symmetric channel. The corresponding decoding radii are
calculated in closed forms and tight bounds on the error probability are derived. The multiplicity assignment
strategy and the corresponding performance analysis are then generalized to characterize the decoding region of
algebraic soft-decision decoding over a mixed error and bit-level erasure channel. The bit-level decoding region
of the proposed multiplicity assignment strategy is shown to be significantly larger than that of conventional
Berlekamp-Massey decoding. As an application, a bit-level generalized minimum distance decoding algorithm is
proposed. The proposed decoding compares favorably with many other Reed-Solomon soft-decision decoding
algorithms over various channels. Moreover, owing to the simplicity of the proposed bit-level generalized
minimum distance decoding, its performance can be tightly bounded using order statistics.
\end{abstract}


\section{Introduction}
\label{sec:introduction}

Reed-Solomon (RS) codes are powerful error correction codes which are widely employed in many state-of-the-art
communication systems. However, in most existing systems, RS codes are decoded via algebraic hard decision
decoding (HDD) which does not fully exploit the error correction capability of the code. Efficiently utilizing
the soft information available at the decoder input to improve the performance of RS codes is a long-time
standing open problem. However, this can be computationally complex and in fact, it has recently been proved
that maximum likelihood decoding (MLD) of RS codes is an NP hard problem \cite{guruswami_np}. Nevertheless, many
efficient suboptimal soft decision decoding (SDD) algorithms have been developed. Some use the reliability value
to assist HDD, \cite{forney_gmd, chase_chase, tang_cga}; some take advantage of the structure of RS codes at the
bit-level to reduce the complexity \cite{vardy_rs, ponn_rs}; some apply decoding algorithms for general linear
block codes to RS soft-decision decoding \cite{hu_osd, fossorier_bma}; and more recently, iterative techniques
have also been proposed for RS soft-decision decoding \cite{ungerboeck_rs, yedidia_gfft, jiang_ssid,
jiang_it06}.

Since the seminal works by Sudan \cite{sudan_sudan}, Guruswami and Sudan \cite{guruswami_gs} and Koetter and
Vardy \cite{koetter_kv}, algebraic soft-decision decoding (ASD) algorithms have gained research interest due to
their significant performance improvement over HDD. From a practical perspective, many techniques have been
proposed to steer ASD decoding towards an implementable alternative to conventional Berlekamp and Massey (BM)
decoding, for example, \cite{gross_kv, gross_vlsi, ahmed_isit, haitao_isita, haitao_thesis, bellorado_kvchase}
and references therein. We also refer interested readers to \cite{mceliece_gsd} for a comprehensive tutorial of
the generic GS algorithm. From a theoretical perspective, however, optimal multiplicity assignment strategy
(MAS) for ASD and corresponding performance analysis still remains an open problem. In \cite{koetter_kv},
Koetter and Vardy presented an asymptotically optimal MAS that maximizes the transmission rate for a given
channel such that the probability of error can be made arbitrarily small as the code length goes to infinity.
Multiplicity assignment optimization for finite length RS codes has been considered in \cite{parv_gauss,
el-khamy_mas, nayak_kv} using numerical algorithms. In \cite{nayak_kv}, MAS for general discrete memoryless channels (DMC) has been investigated and an upper bound based on a Chernoff bounding technique has been derived. However, the Chernoff-type bound \cite{nayak_kv} largely relies on numerical computation and gives little insight into the decoding region of ASD under a certain MAS. Besides, the bound becomes loose for practical high rate RS codes. In \cite{koetter_additive}, a general framework has been studied to optimize the multiplicity assignment for an additive cost function (e.g., Hamming distance, weighted Hamming distance and other error metrics). In fact, as one reviewer pointed out, the optimal decoding radius of ASD over the BEC and the BSC derived in Section \ref{sec:rs_erasure} and Section \ref{sec:rs_bsc} can also be solved using this general framework proposed in \cite{koetter_additive}. However, the multiplicity optimization in \cite{koetter_additive} does not apply to practical noisy channels (such as additive white Gaussian noise (AWGN) channels) in an obvious way. More recently, the decoding region and typical error patterns of ASD with infinite cost over some basic DMC's have been studied independently in \cite{jiang_asd_bec} and \cite{justesen_isit} (which can be viewed as more detailed discussions on some special cases covered by the general framework in \cite{koetter_additive}). Based on the analysis in \cite{jiang_asd_bec}, the performance of practical high rate RS codes can be tightly bounded over erasure channels. However, as suggested in \cite{justesen_isit}, even with infinite cost, ASD has a significant gain over BM only for RS codes of low rates or when the number of candidates at the decoder input is small.

In this paper, we propose a MAS for ASD that provides a significant performance improvement over the BM
algorithm even for high rate RS codes with a computational complexity that is practically affordable. In
contrast to the popular view point that ASD is a symbol-level SDD scheme, we study the performance of ASD of RS
codes using bit-level soft information. We show that carefully incorporating bit-level soft information in
multiplicity assignment and interpolation is the key step to achieve most of the coding gain offered by ASD but
still maintain a moderate complexity. Based on the analysis, a new SDD scheme is proposed for RS codes, which
outperforms many existing ASD algorithms in the literature in terms of both performance and computational
complexity.

The rest of this paper is organized as follows: After a brief review of ASD in Section~\ref{sec:review}, we
investigate optimal MAS's for ASD over erasure channels and binary symmetric channels (BSC) with infinite cost in
Sections~\ref{sec:rs_erasure} and \ref{sec:rs_bsc}. The corresponding decoding region of ASD is characterized
and performance bounds are derived. It is shown that ASD has a significant gain over conventional BM decoding
over binary erasure channels (BEC) and most of the coding gain comes from appropriate multiplicity assignment to
bit-level erasures. On the other hand, the gain of ASD over GS decoding is large only for short length or low
rate RS codes over BSC's. In Section~\ref{sec:bit_region}, the analysis is generalized to a mixed error and
bit-level erasure channel and a simple MAS is proposed. In the infinite cost case, the decoding region of the
proposed MAS is shown to approach the outer bound of optimal MAS for practical medium to high rate RS codes.
In the finite cost case, the decoding region of the proposed MAS is also characterized for any given
multiplicity parameter $M$. By treating erasures at the bit-level, the proposed MAS has a significantly larger decoding region than that of conventional BM and more recent GS algorithm. Based on insights obtained from the
performance analysis, in Section~\ref{sec:bit_gmd}, we develop a sequential MAS called bit-level generalized
minimum distance (BGMD) decoding, which successively erases the least reliable bits (LRB). In spite of its
simplicity, BGMD algorithm provides a significant gain over conventional BM decoding and compares favorably with
many existing MAS's of ASD and other RS SDD schemes over various channels of practical interests. Moreover, due
to its simple structure, the decoding performance of BGMD for practical high rate RS codes can be tightly
bounded using a standard order statistics bounding technique. BGMD upper bound suggests a significant gain
over BM decoding in the high SNR region, where the evaluation of the performance is beyond the capability of
computer simulation but of significant practical value. Simulation results are presented in
Section~\ref{sec:simulation} and conclusion is drawn in Section~\ref{sec:conclusion}.

\section{Reed-Solomon Codes and Algebraic Soft Decision Decoding}
\label{sec:review}

In this section, we review some background materials on RS codes and ASD of RS codes that are relevant to this
paper. Underlined letters will be used to denote vectors and bold face letters will be used to denote matrices
throughout this paper.

\subsection{Evaluation Form of Reed-Solomon Codes}
\label{subsec:evaluation_rs}

Define the message vector $\underline{g}$ as:
\begin{equation} \label{eqn:rs_msg}
\underline{g} = \left(g_0, g_1, \cdots g_{K-1}\right), g_i \in GF(q).
\end{equation}
The polynomial form of the message is:
\begin{equation} \label{eqn:rs_msgpoly}
g(x) = g_0+g_1 x+\cdots+g_{K-1} x^{K-1}
\end{equation}
Let $\Gamma = \{\gamma_1, \gamma_2, \cdots, \gamma_N \}$ be an ordered set of $N$ distinct elements in $GF(q)$. An $(N, K)$ RS codeword is defined by evaluating the message polynomial $g(x)$ at the $N$ distinct points in the ordered set $\Gamma$. The
corresponding RS code is the set of all possible such evaluations:

\begin{equation} \label{eqn:rs_def}
{\mathcal C} (N, K, \{\gamma_1, \gamma_2, \cdots, \gamma_N \}) = \{\left(g(\gamma_1), g(\gamma_2), \cdots, g(\gamma_N)\right)\}
\end{equation}
where $g(x) \in GF_q[x]$ is a polynomial with maximum degree less than $K$.

\subsection{Algebraic Soft-Decision Decoding}
\label{subsec:asd}

Let $\mathcal{A} (X, Y) = \sum_{i = 0}^{\infty}\sum_{j = 0}^{\infty} a_{i, j}X^{i}Y^{j}$ be a bivariate
polynomial over $GF(q)$ and let $w_x$, $w_y$ be nonnegative real numbers. The $(w_x, w_y)$-weighted degree of a
nonzero polynomial $\mathcal{A}(X, Y)$ (denoted as $\texttt{deg}_{w_x, w_y} (\mathcal{A})$) is defined as the
maximum over all numbers $i w_x + j w_y$ such that $a_{i, j} \neq 0$. The $(1, 1)$ degree is usually referred to
as the degree of the polynomial $\mathcal{A} (X, Y)$ (denoted as $\texttt{deg} (\mathcal{A})$). The bivariate
polynomial $\mathcal{A}(X, Y)$ is said to pass through a point $(\alpha, \beta)$ with multiplicity $m$ (or
equivalently, the point $(\alpha, \beta)$ is said to be a zero of multiplicity $m$ of the polynomial
$\mathcal{A}(X, Y)$), if $\mathcal{A}(X+\alpha, Y+\beta)$ contains a monomial of degree $m$ and does not contain
any monomials of degree less than $m$.

Suppose an RS codeword is modulated and transmitted through a memoryless channel. Let $\mathcal{\underline{X}} = (\mathcal{X}_1, \mathcal{X}_2, \cdots, \mathcal{X}_N)$ be the random variable of the transmitted codeword and $\underline{x} = (x_1, x_2, \cdots, x_N)$ be the realization of the random variable. Let $\underline{\mathcal{Y}} =
(\mathcal{Y}_1, \mathcal{Y}_2, \cdots, \mathcal{Y}_N)$ be the random variable of the channel output and $\underline{y} =
(y_1, y_2, \cdots, y_N$) a particular realization of the channel output. The {\em a posteriori} probability
(APP) of each symbol without considering the constraint imposed by the RS code can be computed from the channel output as:
\begin{equation}
\label{eqn:rs_app}
\pi_{i, j} = Pr\left(\mathcal{X}_j = \alpha_i| \mathcal{Y}_j = y_j\right), 1 \le i \le q, 1 \le j \le N
\end{equation}
where $\{\alpha_1, \alpha_2, \cdots, \alpha_q\}$ are all distinct elements in $GF(q)$. Define the $q \times N$
reliability matrix $\mathbf{\Pi}$ as a matrix with entries $\{\pi_{i, j}\}$ as computed in (\ref{eqn:rs_app}).
$\mathbf{\Pi}$ serves as a soft input to an ASD decoder. The generic algorithm of ASD as in \cite{koetter_kv} is
described in the following 4 steps:

\emph{\underline{Multiplicity Assignment}}: Compute the multiplicity matrix $\textbf{M}$ with integer entries
$\{M_{i,j}\}$ based on the reliability matrix $\mathbf{\Pi}$ according to a particular multiplicity assignment strategy.

\emph{\underline{Interpolation}}: For a given multiplicity matrix $\textbf{M}$ with entries $\{M_{i,j}\}$,
construct a bivariate polynomial $\mathcal{A}(X, Y)$ of minimal $(1, K-1)$-weighted degree that passes through
each point $(\gamma_j, \alpha_i)$ with multiplicity at least $M_{i,j}$, for $i = 1, 2, \cdots, q$ and $j = 1, 2,
\cdots, N$.

\emph{\underline{Factorization}}: Find all polynomials $g(X)$ such
that $(Y-g(X))$ divides $\mathcal{A}(X, Y)$ and
$\texttt{deg}(g(X)) < K$. Form a candidate codeword list by
re-encoding all such polynomials $g(X)$.

\emph{\underline{Codeword Selection}}: If the candidate codeword list is not empty, select the most likely
codeword from the list as the decoder output; otherwise, declare a decoding failure.

Intuitively, the idea of ASD is to take advantage of soft information and assign higher multiplicities to more
probable symbols such that the decoder has a better chance to find the correct codeword.

\subsection{Performance of Algebraic Soft-decision Decoding}
\label{subsec:perf_asd}

Define the inner product between two $q \times N$ matrices as:
\begin{equation} \label{eqn:rs_innprod}
\langle \textbf{A}, \textbf{B}  \rangle \mathop{=}^{\texttt{def}}\texttt{trace}(\textbf{A} \textbf{B}^{T}) =
\sum_{i = 1}^{q}\sum_{j = 1}^{N}a_{i,j}b_{i,j}
\end{equation}

Let \textbf{1} be the all-one $q \times N$ matrix. Suppose the vector $\underline{x}$ represents an RS
codeword, let $\left[\underline{x}\right]$ be the codeword matrix with entries
$\left[\underline{x}\right]_{i,j}$ defined as: $\left[\underline{x}\right]_{i,j} = 1$ if
${x}_{j} = \alpha_{i}$; $\left[\underline{x}\right]_{i,j} = 0$, otherwise. As in
\cite{koetter_kv}, the score and cost are defined as follows.

\begin{defn}
The score $S_{\textbf{\textsc{M}}}(\underline{x})$ with respect to a codeword $\underline{x}$ for a given multiplicity matrix $\textbf{\textsc{M}}$ is defined as:
\begin{equation} \label{eqn:score_org} S_{\textbf{\textsc{M}}}(\underline{x}) = \langle \textbf{\textsc{M}},
\left[\underline{x}\right]\rangle
\end{equation}
\end{defn}

\begin{defn} The cost $C_{\textbf{\textsc{M}}}$ of a given multiplicity matrix $\textbf{\textsc{M}}$ is defined as the number of linear constraints imposed by $\textbf{\textsc{M}}$:
\begin{equation} \label{eqn:cost_org}
C_{\textbf{\textsc{M}}} = \frac{1}{2}\sum_{i=1}^{q}\sum_{j=1}^{N} M_{i,j}(M_{i,j}+1) = \langle \textbf{\textsc{M}}, \textbf{\textsc{M+1}} \rangle/2
\end{equation}
\end{defn}
It is understood that the score and the cost are associated with a particular multiplicity matrix $\textbf{M}$. Besides, unless otherwise specified, the score is taken with respect to the transmitted codeword. Therefore, for brevity, we use $S$ to denote $S_{\textbf{M}}(\underline{x})$ and $C$ to denote $C_{\textbf{M}}$ throughout the rest of this paper.

Similar to other list decoding algorithms, the probability of error of ASD can be upper bounded using the union bound:
\begin{equation}
\label{eqn:asd_ub} P_{ASD} \le P_{List} + P_{ML}
\end{equation}
where $P_{List}$ is the probability that the transmitted codeword is not on the list and $P_{ML}$ is the
probability that the maximum likelihood decision is not the transmitted codeword. It is known in the literature (see \cite{jiang_it06} \cite{koetter_kv} \cite{gross_kv}) that $P_{List} \gg P_{ML}$ for many practical high rate long RS codes. Therefore, we approximate $P_{ASD} \approx P_{List}$ throughout the rest of this paper (see also \cite{nayak_kv} \cite{justesen_isit}). However, for low rate codes, $P_{ML}$ must be computed to evaluate $P_{ASD}$ (see \cite{duggan_mlbound} for details). In general, the decoding region of ASD is difficult to characterize and analytical computation of $P_{List}$ is a tough problem. However, it is shown in \cite{koetter_kv} that the transmitted codeword is guaranteed to be on the list when the following sufficient condition is satisfied:

\begin{lem}\label{thm:guaranteed_decoding_org} \cite{koetter_kv}
Finite cost: A sufficient condition for the transmitted codeword to be in the list is:
\begin{equation}
\label{eqn:sufficient_org} S > \min \left\{ \delta \in \mathbb{Z}: \left\lceil \frac{\delta+1}{K-1}\right\rceil
\left(\delta+1-\frac{(K-1)}{2}\left\lfloor \frac{\delta}{K-1}\right\rfloor \right) > C \right\}
\end{equation}
\end{lem}
The proof of Lemma \ref{thm:guaranteed_decoding_org} is given in Theorem 3 in \cite{koetter_kv}. The above
sufficient condition can also be expressed as (see also \cite{mceliece_gsd}): the transmitted codeword will be
in the list if
\begin{eqnarray}
\label{eqn:sufficient_form2}  T(S) &>& C \\
\nonumber  \text{where~} T(S) &=& (a+1)\left[S-\frac{a}{2}(K-1)\right], a(K-1) < S \le (a+1)(K-1), a = 0, 1, \cdots
\end{eqnarray}
is a piecewise linear function.

Generally speaking, larger cost leads to better decoding
performance, while it also increases complexity (though the
performance of ASD does not monotonically improve as the cost
increases). As the cost goes to infinity, we can further simplify
the sufficient condition as:
\begin{lem}\label{thm:guaranteed_decoding}\cite{koetter_kv}
Infinite cost: A sufficient condition for the transmitted codeword to be in the list as $C \rightarrow \infty$ is:
\begin{equation}
\label{eqn:sufficient} S \ge \sqrt{2(K-1)C}
\end{equation}
\end{lem}
See Corollary 5 in \cite{koetter_kv} for the proof.

Usually, the sufficient conditions (\ref{eqn:sufficient_org}) and (\ref{eqn:sufficient}) become tight
approximations when $N$ is large (see \cite{koetter_kv}, \cite{vardy_notes}). With a little bit abuse of terminology, we have the following definition:
\begin{defn}
\label{defn:failure1} For ASD with finite cost, a received word is said to be certainly decodable if and only if the
sufficient condition (\ref{eqn:sufficient_org}) is satisfied.
\end{defn}
\begin{defn}
\label{defn:failure2} For ASD with infinite cost, a received word is said to be certainly decodable if and only if the
sufficient condition (\ref{eqn:sufficient}) is satisfied.
\end{defn}
For the rest of the paper, we approximate the actual decoding
error probability of ASD by the probability that a codeword is \emph{NOT} certainly decodable. An ASD error is declared when the received word is not certainly decodable as defined in Definition \ref{defn:failure1} and Definition \ref{defn:failure2} for finite cost and infinite cost cases respectively. Practically speaking, though the decoder may still be able to list the transmitted codeword even when the sufficient condition is not satisfied, the probability is very small and the approximation is tight for long codes, which are used in many existing standards (see \cite{koetter_kv}, \cite{gross_kv} and \cite{vardy_notes}).

\subsection{Binary Image Expansion of Reed-Solomon Codes}
\label{subsec:perf_asd}

RS codes can also be viewed as non-binary Bose-Chaudhuri and Hocquenghem (BCH) codes. The parity-check matrix $\textbf{H}_s$ of a narrow sense RS$(N, K)$ code over GF($q$) with a minimum distance $d_{min} = N-K+1$ can be represented by:

\begin{equation} \label{eqn:prt_matrix}
   \textbf{H}_{s}=\left(\begin {array}{cccccc}
                     1 &  \alpha   & \cdots & \alpha^{(N-1)}\\
                     1 &  \alpha^2 & \cdots & \alpha^{2(N-1)}\\
                       &          & \cdots &        \\
                     1 &  \alpha^{(N-K)} & \cdots &
\alpha^{(N-K)(N-1)}\\
              \end{array} \right)\;
\end{equation}
where $\alpha$ is a primitive element in $GF(q)$.

RS codes over $GF(2^m)$ (which are of most practical interest) can
be represented using equivalent binary image expansions
\cite{lin_book}. Let $n = N \times m$ and $k = K \times m$ be the
length of the codeword and the message at the bit-level,
respectively. $\textbf{H}_{s}$ has an equivalent binary image
expansion $\textbf{H}_{b}$ (see \cite{lin_book} for details),
where $\textbf{H}_{b}$ is an $(n-k) \times n$ binary parity check
matrix. In other words, an RS code can also be viewed as a binary
linear block code. In many practical systems, RS codewords are
mapped into their binary image expansions and then transmitted
through channels using binary modulation formats. Therefore, it is of both theoretical and practical value to study ASD using
bit-level soft information.

\section{Performance Analysis of Algebraic Soft-decision Decoding over Erasure Channels}
\label{sec:rs_erasure}

In this section, we consider MAS's for erasure channels and their corresponding performance analyses.

\subsection{Algebraic Soft-decision Decoding over the Binary Erasure Channel}
\label{subsec:rs_bec}

We first consider the case when RS codewords are transmitted as bits through a BEC with erasure probability $\epsilon$.  Note that we are interested in both the truly optimal multiplicity assignment, (namely, for a given cost, the score is maximized) and the asymptotically optimal multiplicity assignment that maximizes the number of correctable bit-level erasure in the worst case (which can also be solved from the general formula in \cite{koetter_additive}). However, it should be noted that the MAS we are considering here does not take into account the inherent structure of RS code, which is extremely challenging \cite{koetter_kv}. As a result, similar to the ``adversarial channel optimal''argument in \cite{koetter_kv}, \cite{nayak_kv}, \cite{el-khamy_mas}, we assume that the symbols in a codeword are independent and identically distributed (i.i.d) with a uniform distribution over $GF(q)$ during the multiplicity assignment stage.
Along the way of these derivations, we also gain some insight on the dominating erasure patterns over the BEC.

We first define the ``candidate'' of a symbol and the ``type'' of a symbol as follows:
\begin{defn}
\label{defn:candidate}
A binary m-tuple $\alpha_i \in \{0, 1\}^m$ is a $candidate$ for a received m-tuple symbol $y_j \in \{0, 1, \epsilon \}^m$ if each non-erased bit of $y_j$ agrees with the corresponding bit of $\alpha_i$.
\end{defn}

\begin{defn}
\label{defn:bec_type}
The $type$ of a received m-tuple symbol $y_j \in \{0, 1, \epsilon\}^m$ over the BEC is the number of bit-level erasures in $y_j$.
\end{defn}

For BEC's, according to the ``optimal for the adversarial channel model'' argument, a natural MAS is to assign equal multiplicity to each candidate in the same symbol. Consequently, there are $2^i$ candidates for a symbol of type $i$. For a codeword over $GF(2^m)$, there are at most $(m+1)$ types of symbols. Let $a_i$ be the number of symbols of type $i$ in a received word. As discussed above, we will assign the same multiplicity to each candidate in the same symbol. Moreover, we assume the same multiplicities are assigned to all candidates of the same type in a received word; whereas, the multiplicity assigned to type $i$ may vary according to the received word. This assumption will be justified later. Let $m_i$ be the multiplicity assigned to each candidate of type $i$. Thus, the total multiplicity
assigned to one symbol of type $i$ is $2^i m_i$. The score (with respect to the transmitted codeword) and cost are:

\begin{eqnarray}
\label{eqn:score_defn} S & = & \sum_{i=0}^{m}{a_i m_i} \\
\label{eqn:cost_defn} C & = & \sum_{i=0}^{m}{a_i 2^i {{m_i+1}\choose{2}}}\\
\label{eqn:cost_defn2} C & = & \frac{1}{2}\sum_{i=0}^{m}{a_i 2^i m_i^2(1+o(1))}, m_i \rightarrow \infty
\end{eqnarray}

The approximation in (\ref{eqn:cost_defn2}) becomes tight when $m_i$ becomes large. We will derive an upper
bound and a lower bound on the probability of ASD error with infinite cost as defined in
Definition \ref{defn:failure2}. Furthermore, we consider ASD with infinite cost such that we can relax the
multiplicities from being integers to real numbers. It is justified by the fact that rational numbers are dense
on the real axis and every finite collection of rational numbers can be scaled up to integers if there is no cost constraint. Hence any real number multiplicity assignment scheme can be approximated arbitrarily close with integer number multiplicity assignment scheme (see also \cite{el-khamy_mas}).

Following \cite{koetter_kv} \cite{gross_kv}, we define proportional multiplicity assignment strategy (PMAS) as
follows:

\emph{\underline{Proportional Multiplicity Assignment Strategy}}: For a given total multiplicity per symbol
$M$, PMAS assigns multiplicity proportional to the APP of that candidate, i.e., the multiplicity
we assign to $\alpha_k$ in the $j^{th}$ symbol of the received vector is $M_{k, j} = \lfloor
\pi_{k, j} M \rfloor$, where $M$ is a predetermined real number.

For example, in the case of BEC, $M_{k, j} = \lfloor \pi_{k, j} M \rfloor = \lfloor 2^{-i} M \rfloor$ if $\alpha_k$ is a candidate for the $j^{th}$ symbol, which is a type $i$ symbol; $M_{k, j} = 0$ if $\alpha_k$ cannot be a candidate for the $j^{th}$ symbol. PMAS defined above is equivalent to the simplified KV defined in \cite{gross_kv}. Note that there is a quantization error, however, the multiplicity assignment is asymptotically proportional to the APP as the cost approaches infinity. We will show in the following that PMAS is optimal over the BEC with infinite cost:

\begin{thm}\label{thm:proportional_assignment}
The proportional multiplicity assignment strategy (PMAS) is optimal over the BEC regardless of the received
signal, i.e., PMAS maximizes the score for a given cost over the BEC.
\end{thm}
\begin{proof} \label{prf:proportional_assignment}
Assume that the received word has $a_i$ symbols of type $i$, the MAS can be formulated as maximizing the score with a cost constraint. With infinite cost, the problem is expressed as:
\begin{eqnarray}
   \max_{\{m_i\}}~~~S &=& \sum_{i=0}^{m}{a_im_i} \\
\nonumber   \text{subject to~~~} C &\approx& \frac{1}{2}\sum_{i=0}^{m}{a_i 2^i m_i^2}\le C_0
\end{eqnarray}
This is a standard optimization problem with linear cost function and quadratic constraint. Using a Lagrange
multiplier, the new objective function becomes
\begin{equation}
\label{eqn:Lagrange} \emph{L} = -\sum_{i=0}^{m}{a_im_i}+\lambda \left(\frac{1}{2}\sum_{i=0}^{m}{2^ia_im_i^2}-C_0
\right)
\end{equation}

Take the partial derivative with respect to $m_i$ and set it to zero. We have:
\begin{equation}
\label{eqn:Lagrange partial} \frac{\partial{\emph{L}}}{\partial{m_i}} = -a_i+\lambda 2^ia_im_i = 0
\end{equation}

Therefore we have $m_i = \frac{2^{-i}}{\lambda}$, i.e., $m_i \propto 2^{-i}$, which proves that PMAS is optimal.
\end{proof}
Note that $m_i$ does not depend on $a_i$. Even without the assumption that equal multiplicities are assigned to
candidates of the same type, we still get $m_i = \frac{2^{-i}}{\lambda}$ for all type $i$ candidates,
i.e., PMAS is optimal over the BEC.

Since PMAS is optimal over the BEC, we will from now on assume that PMAS is used. Under PMAS, we assume that the
total multiplicity for each symbol is $M$. Consequently, the score is $S_0 = \sum_{i=0}^{m} a_i 2^{-i}
M = \eta M$ and the cost is $C_0 = \frac{1}{2}\sum_{i=0}^{m} a_i 2^{-i} M^2(1+o(1)) =
\frac{1}{2} \eta M^2(1+o(1))$, where $\eta = \sum_{i=0}^{m} a_i 2^{-i}$ is a positive number. The sufficient condition of (\ref{eqn:sufficient}) becomes:
\begin{align}
\label{align:sufficient_0}
S_0 &\ge \sqrt{2(K-1)C_0}\\
\label{align:eta_ieq}\eta &> K-1
\end{align}
When $K = 1$, under PMAS, $\eta > 0$, the transmitted codeword will always be on the decoding list. From now on,
we only consider the case $K > 1$.

We study the worst case bit-level decoding radius of ASD under PMAS with infinite cost over the BEC. We need the
following lemmas.
\begin{lem}\label{lem:monotone erasure}
Over the BEC, if a received word $\underline{y}$ is certainly decodable under PMAS with infinite cost, then any other word $\underline{y}'$, in which only a subset of the erased bits in $\underline{y}$ are erased, is always certainly decodable.
\end{lem}
\begin{proof} \label{prf:monotone_erasure}
The proof is immediate by the fact that if some of the erasures are recovered, $\eta$ will increase and as can
be seen from (\ref{align:eta_ieq}), the decoding performance monotonically improves as $\eta$ increases.
\end{proof}

\begin{lem}\label{lem:worst case erasure pattern}
Over the BEC, given $f$ bit erasures, the worst case erasure pattern for ASD under PMAS with infinite cost is
that all erased bits are spread in different symbols as evenly as possible. That is: $(N - f + \lfloor \frac{f}{N}
\rfloor N)$ symbols contain $\lfloor \frac{f}{N} \rfloor$ bit erasures and $(f - \lfloor \frac{f}{N} \rfloor N)$
contain $\lceil \frac{f}{N} \rceil$ bit erasures.
\end{lem}
\begin{proof} \label{prf:worst case erasure pattern}
Take two arbitrary symbols of type $i$ and $j$, if we average the bit erasures between these two, we get two
symbols of type $\lfloor \frac{i+j}{2} \rfloor$ and $\lceil \frac{i+j}{2} \rceil$. The updated $\eta'$ can be
expressed as:
\begin{equation}\label{eqn_eta}
\eta' = \eta + 2^{-\lfloor \frac{i+j}{2} \rfloor} + 2^{-\lceil \frac{i+j}{2}\rceil} - 2^{-i} - 2^{-j} \le \eta
\end{equation}
Since $\eta \ge \eta'$ and again according to (\ref{align:eta_ieq}), the latter erasure pattern is worse. By
repeating the above procedure, we can finally get the worst case erasure pattern of ASD under PMAS, i.e., the
bit erasures are spread as evenly as possible in different symbols.
\end{proof}

According to Lemma~\ref{lem:monotone erasure} and Lemma~\ref{lem:worst case erasure pattern}, the bit-level
decoding radius of PMAS can be characterized.
\begin{thm}\label{thm:worst_region_bec}
Over the BEC, every pattern of f bit erasures can be corrected with ASD under PMAS with infinite cost if the following is satisfied:
\begin{eqnarray}
\label{eqn:bec_radius_2r} f &<& (i+1)N-2^i(K-1), \text{~for~} 2^{-i}+\frac{1-2^{-(i+1)}}{N} \le R <
2^{-(i-1)}+\frac{1-2^{-i}}{N}, i = 1, 2, \cdots, m
\end{eqnarray}
Especially, for high rate RS codes, we have:
\begin{eqnarray}
\label{align:opt_e_a} f &<& 2N-2(K-1), \text{~for~} R \ge \frac{1}{2}+\frac{3}{4N}
\end{eqnarray}

\end{thm}
\begin{proof} \label{prf:worst_region_bec}
According to Lemma \ref{lem:worst case erasure pattern}, the worst case erasure pattern is all erased bits spread evenly over different symbols. First consider the case $f \le N$, (\ref{align:eta_ieq}) becomes:
\begin{align}
(N-f)+\frac{1}{2}f > K-1
\end{align}
The corresponding rate region must satisfy the constraint that when $f = N+1$, in the worst case $\eta \le K-1$.
We get $K \ge \frac{1}{2}N+\frac{3}{4}$ in this case. Altogether, we get the bit-level decoding radius for the
high rate case:
\begin{align}
f < 2N-2(K-1), \text{~when~} R \ge \frac{1}{2}+\frac{3}{4N}
\end{align}
Similarly, when the decoding radius $(i-1)N < f \le iN$, we must have $2^{-i}N+1-2^{-(i+1)} \le K <
2^{-(i-1)}N+1-2^{-i}$, where $i = 1, 2, \cdots, m$. We can obtain the exact decoding radius for all these cases:
\begin{align}
\label{eqn:bec_radius_final}f < (i+1)N-2^i(K-1), \text{~when~} 2^{-i}N+1-2^{-(i+1)} \le K < 2^{-(i-1)}N+1-2^{-i}
\end{align}
\end{proof}
According to Theorem~\ref{thm:worst_region_bec}, any erasure pattern with $f < 2(N-K+1)$ is certainly decodable. We can
get an upper bound on the frame error rate (FER) of ASD under PMAS with infinite cost.

\begin{cor}\label{cor:lower_bound}
For RS codes of rate $R \ge \frac{1}{2}+\frac{3}{4N}$ over the BEC, a received vector is NOT certainly decodable under PMAS with infinite cost when there are more than $2(N-K)+1$ symbols containing erased bits.
\end{cor}
\begin{proof} \label{prf:lower_bound}
The corollary follows from (\ref{align:opt_e_a}) and Lemma \ref{lem:monotone erasure}. If there are more than
$2(N-K)+1$ symbols having erased bits, the most optimistic case is that these symbols are of type $1$. Besides,
due to (\ref{align:opt_e_a}), the sufficient condition is not satisfied as defined in Definition
\ref{defn:failure2}.
\end{proof}

\begin{figure}
\begin{center}
\includegraphics[width=4.0in]{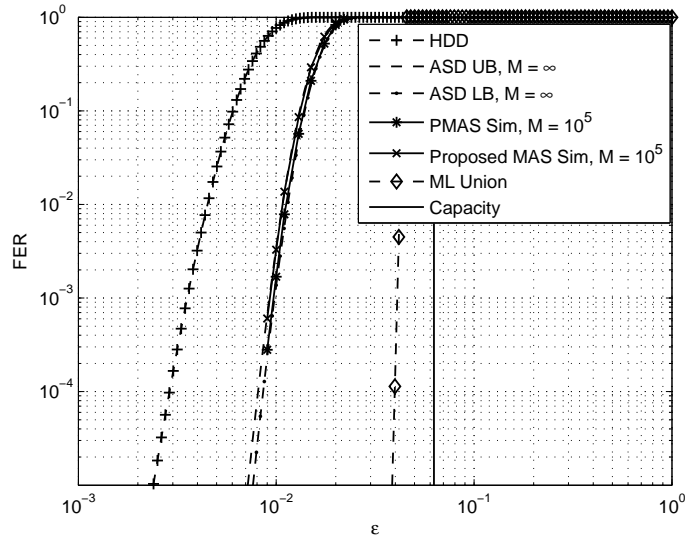}
\caption{Bounds and Simulation Results of ASD for RS(255,239) over the BEC} \label{fig:upp_low_bd}
\end{center}
\end{figure}

Theorem \ref{thm:worst_region_bec} gives an upper bound on the FER performance over the BEC and Corollary
\ref{cor:lower_bound} provides a lower bound. These bounds are shown in Figure \ref{fig:upp_low_bd} in
conjunction with the union bound on the averaged FER of a maximum likelihood (ML) decoder over the RS ensemble
\cite{retter_gs}. Note that ASD has a significant performance gain over conventional BM erasure decoding. This
is intuitive since we do not have to erase the whole symbol if some bits in the symbol are erased, which can be
taken advantage of by ASD. It can be seen from the figure that for practical high rate long codes, both the
upper and lower bounds are tight and they together accurately indicate the performance of ASD. Also note that
the averaged performance of the ML decoder over the RS ensemble is very close to the capacity of the BEC, which
shows that RS codes are good codes.

The tightness of the upper bound and the lower bound of ASD motivates the following proposed MAS:

\emph{\underline{Proposed Multiplicity Assignment Strategy}}: In each received symbol, we assign $m_0 =
M$ if that symbol does not contain erased bits, assign $m_1 = M/2$ if the symbol contains 1
bit-level erasure and assign zero multiplicity for symbols containing more than 1 bit-level erasures, that is to
set $m_j = 0, j = 2, \cdots, m$.

\begin{rem}
\label{rem:bitmas} Since erasing 2 bits in the same symbol leads to the same score but less cost than 2 bits in
two different symbols,the worst case erasure pattern of the proposed MAS for RS codes with $R \ge
\frac{1}{2}+\frac{3}{4N}$ is that all bit-level erasures are spread in different symbols. According to
Theorem~\ref{thm:worst_region_bec}, the proposed MAS can recover any bit-level erasures containing less than
$2(N-K+1)$ bit erasures. Essentially, the proposed MAS takes care of the worst case erasure pattern only and it
is asymptotically optimal in terms of achieving the largest worst case decoding radius. Consequently, the FER
upper bound derived in Theorem~\ref{thm:worst_region_bec} for $R \ge \frac{1}{2}+\frac{3}{4N}$ is also a valid
upper bound for the proposed MAS. Though the proposed MAS is not optimal as PMAS, the loss is quite small by
comparing the upper bound of the proposed MAS and the lower bound of PMAS for high rate RS codes. It can also be
seen from Figure~\ref{fig:upp_low_bd} that the simulation results of the proposed MAS and optimal PMAS are
very close.
\end{rem}

\subsection{Extension to $2^u$-ary Erasure Channels}
\label{subsec:rs_2^u-ary}

We extend the result in Subsection \ref{subsec:rs_bec} to $2^u$-ary erasure channels, i.e. $u$ coded bits are
grouped together and transmitted using a $2^u$-ary symbol (it can be QAM or PSK modulation format). The channel
will erase the signal with erasure probability $\epsilon$ at $2^u$-ary symbol-level. Practical channels of this
model were discussed in \cite{wesel_robust}.

In Subsection \ref{subsec:rs_bec}, we showed that PMAS is optimal for erasure channels. Clearly, all erasure
patterns in this $2^{u}$-ary erasure channel model is a subset of erasure patterns of BEC. Therefore, with
infinite cost, PMAS is still optimal for this channel model. Here, we only consider the case when $u$ divides
$m$, i.e., $m = vu$. Thus, for each symbol, we have $(v+1)$ types.

\begin{lem}\label{thm:worst case modulation}
Over the $2^u$-ary erasure channel, the worst case erasure pattern for ASD under PMAS with infinite cost is that
all erasure events are spread in different symbols as evenly as possible.
\end{lem}

\begin{proof} \label{prf:worst case modulation}
Assume two RS symbols are of type $i$ and $j$, we can average the erasure events between the two symbols, we
have:
\begin{align}
\label{align:eta_rs_mod} \eta' = \eta +2^{-\lfloor \frac{i+j}{2} \rfloor u}+2^{-\lceil \frac{i+j}{2} \rceil
u}-2^{-iu}-2^{-ju} \le \eta
\end{align}
Similar to Lemma \ref{lem:worst case erasure pattern}, spreading erasure events in different RS symbols evenly
gives the worst case.
\end{proof}

\begin{thm}\label{thm:bounds_memory}
Over the $2^u$-ary erasure channel, ASD under PMAS can guarantee to decode up to $f < (N-K+1)/(1-2^{-u})$
$2^u$-ary symbol-level erasures if $R \ge 2^{-u}+\frac{1+2^{-2u}-2^{-u}}{N})$.
\end{thm}
\begin{proof} \label{prf:bounds_memory}
According to Lemma \ref{thm:worst case modulation} and (\ref{align:eta_ieq}), spreading erasure events in
different symbols is the worst case erasure pattern if $K \ge 2^{-u}N+1+2^{-2u}-2^{-u}$, that is $R \ge
2^{-u}+\frac{1+2^{-2u}-2^{-u}}{N}$. Thus $\eta = N-(1-2^{-u})f$. According to (\ref{align:eta_ieq}) when the
following condition is satisfied:
\begin{align}
\label{align:region}  f < (N-K+1)/(1-2^{-u})
\end{align}
the transmitted codeword is guaranteed to be on the list.
\end{proof}

\begin{rem}
Note that (\ref{align:region}) is a generalization of Theorem \ref{thm:worst_region_bec} (with $u = 1$ as a
special case). As u becomes larger, the asymptotical $2^u$-ary symbol-level decoding radius gets smaller. As $u
\rightarrow m$, $f$ gets close to the conventional BM erasure decoding region.
\end{rem}

\section{Performance Analysis of ASD over the BSC}
\label{sec:rs_bsc}

In this section, we study the performance of ASD over BSC's. For BSC's, both the transmitted and the received
symbols are in $GF(q)$, i.e., $x_j \in GF(q)$ and $y_j \in GF(q)$, for $j = 1, \cdots, N$.
In this case, bit-level reliabilities are the same for all received bits. However, symbol-level soft information
can be utilized by ASD under proper MAS. In the BSC case, the ``type'' of a symbol is related to its bit-level Hamming distance between $x_j$ and $y_j$.
\begin{defn}
\label{defn:bsc_type}
The $type$ of a received m-tuple symbol $y_j \in \{0, 1\}^m$ over the BSC is defined as the bit-level Hamming distance between $x_j$ and $y_j$.
\end{defn}
Let $a_i$ denote the number of symbols of type $i$.
Again, the same multiplicity is assigned to candidates of the same type, that is, we set $M_{k,j} = m_{H(k,j)}$, where $H(k,j)$ is the Hamming distance between symbol $\alpha_k$ and the received symbol $y_j$. For example, we assign $m_0$ to the received symbol, $m_1$ to all the $m$ symbols which differ from the received
symbol in 1-bit position and so on. Note that $M_{k, j}$ is the multiplicity assigned to $\alpha_k$ in the $j^{th}$ symbol and $m_{i}$ is the predetermined multiplicity we will assign to candidates of type $i$. However, unlike the BEC case, the type of the transmitted symbol is unknown at the receiver, the truly optimal MAS is not easy to derive as in the BEC case. Therefore, we resort to the asymptotically optimal MAS, i.e., maximizing the bit-level decoding radius for the worst case error pattern. It can also be easily justified that non-uniform multiplicity assignment among candidates of the same type is strictly suboptimal in terms of achieving the largest decoding radius for the worst case error pattern, since the worst case error pattern will always correspond to the candidates with the smallest multiplicities. The MAS optimization problem can be formulated as a max-min problem over $\{a_i\}$ and $\{m_i\}$.

\begin{eqnarray}
\label{eqn:min_max}
\max_{\{m_i\}}\min_{\{a_i\}} e &=& \sum_{i=0}^{m}{i a_i} - 1 \\
\text{s.~t.~} \sum_{i=0}^{m}{a_i m_i} \nonumber&\le& \sqrt{2(K-1)N\sum_{i=0}^{m}{{{m}\choose{i}}\frac{m_i^2}{2}}}\\
\sum_{i=0}^{m}{a_i} \nonumber &=& N, \text{where~} a_i \text{~are non-negative integers}
\end{eqnarray}
where $e = \sum_{i=0}^{m}{i a_i} - 1$ is the number of bit errors, $S = \sum_{i=0}^{m}{a_i m_i}$ is the score, $C = \sum_{i=0}^{m}{{{m}\choose{i}}\frac{m_i^2}{2}}$ is the cost (infinite cost approximation to be more exact) over the BSC and $\sum_{i=0}^{m}{a_i}$ is the total number of symbols of the RS codeword. It can be seen that (\ref{eqn:min_max}) maximizes the bit-level error decoding radius of ASD in the worst case.

Note that the optimization problem in (\ref{eqn:min_max}) is subsumed by the general formula in \cite{koetter_additive}, as the bit-level errors can also be viewed as an additive cost. In this paper, we take a different approach to derive the optimal decoding radius (another similar derivation can be found in \cite{justesen_isit}). We first take one step back and consider a special case of BSC, called 1-bit flipped BSC, i.e., in each symbol, at most 1 bit is in error. By doing that, we only have to assign multiplicities to two types of candidates. An optimal MAS for this 1-bit flipped BSC is to assign $M$ to $y_{j}$'s and $t M$ to all their 1-bit flipped neighbors. The asymptotically optimal decoding radius $e_{max}$ and the corresponding optimal $t$ can be
computed in close forms. The derivations are given in Appendix \ref{pdx:1bit_bsc}.

It should be noted that for the 1-bit flipped BSC, the performance improvement of ASD over GS is significant
only when the rate is low. For instance, for $N = 255$, $K = 223$, ASD does not increase the bit-level error
correcting radius, for $N = 255$, $K = 167$, ASD gives an extra error correction capability over GS decoding at
the bit-level, for $N = 255$, $K = 77$, it corrects 7 more errors and for $N = 255$, $K = 30$, it corrects 45
more errors. For $K < 30$, all errors can be corrected for the 1-bit flipped BSC.

Now, we show that the above proposed MAS is also asymptotically optimal for RS codes over the BSC under certain
conditions, which are satisfied for a wide range of code rates. We begin with the following Lemma.

\begin{lem}\label{thm:worst case error pattern}
Over the BSC, the worst case error pattern for the proposed MAS with infinite cost is all erroneous bits spread
in different symbols, if the total number of bit errors $e\le N$ and the optimal multiplicity coefficient $t \le
\frac{1}{2}$.
\end{lem}

\begin{proof}\label{prf:worst case error pattern}
Assume $e \le N$ bits get flipped by the channel. If bit errors are spread in different symbols, as the 1-bit
flipped BSC case, the score can be expressed as:

\begin{equation}
\label{eqn:org_score} S = M[(N-e)+t e]
\end{equation}
The cost of ASD for the BSC does not change when the MAS is fixed. For a given number of bit errors, the worst case error pattern minimizes the score of the MAS. In the above MAS, multiplicities are assigned only to $y_{j}$'s and their 1-bit flipped neighbors. Thus, a potentially worse error pattern than the 1-bit flipped BSC is to group bits in some 1-bit-flipped symbols to reduce the score.

Let the worst case error pattern have $e'$ symbols containing 1-bit error and $e''$ symbols containing more than
1-bit errors. Evidently, for symbols containing more than 2-bit errors, we can always further decrease the score
of the proposed MAS for 1-bit flipped BSC by splitting these bit errors into one symbol containing 2-bit errors
and the other containing the remaining errors. Consequently, the worst case error pattern will contain symbols
with at most 2-bit errors. We have $e'+2e'' = e$. The score becomes:

\begin{equation}
\label{eqn:2bit_score} S'' = M[(N-e'-e'') + t e'] = M[(N-e) + e t + e'' (1-2t)]
\end{equation}
When $t \le \frac{1}{2}$, $S'' \ge S$, which proves that spreading all erroneous bits in different symbols is
the worst case error pattern for the proposed MAS over the BSC.
\end{proof}

\begin{thm}\label{thm:bsc_bound}
In the infinite cost case, the proposed optimal MAS for the 1-bit flipped BSC is also optimal for the BSC if the optimal decoding radius of ASD over the 1-bit flipped BSC $e_{max} \le N$ (as given in (\ref{eqn:slope solution})) and the corresponding optimal multiplicity coefficient $t \le \frac{1}{2}$. The optimal bit-level decoding radius of ASD over the BSC is also $e_{max}$.
\end{thm}
\begin{proof} \label{prf:bsc_bound}
According to Lemma \ref{thm:worst case error pattern}, over the BSC, all erroneous bits spread in different
symbols is the worst case error pattern for the proposed MAS  if $e \le N$ and $t \le \frac{1}{2}$, which is
nothing but the 1-bit flipped BSC's. On the other hand, the proposed MAS derived in Appendix~\ref{pdx:1bit_bsc}
is asymptotically optimal for the 1-bit flipped BSC, i.e., maximizing the worst case decoding radius $e_{max}$.
Consequently, the proposed MAS will guarantee to decode all error patterns with no more than $e_{max}$-bit
errors over BSC's as well.
\end{proof}

\begin{figure}[h]
\begin{center}
\includegraphics[width=4.0in]{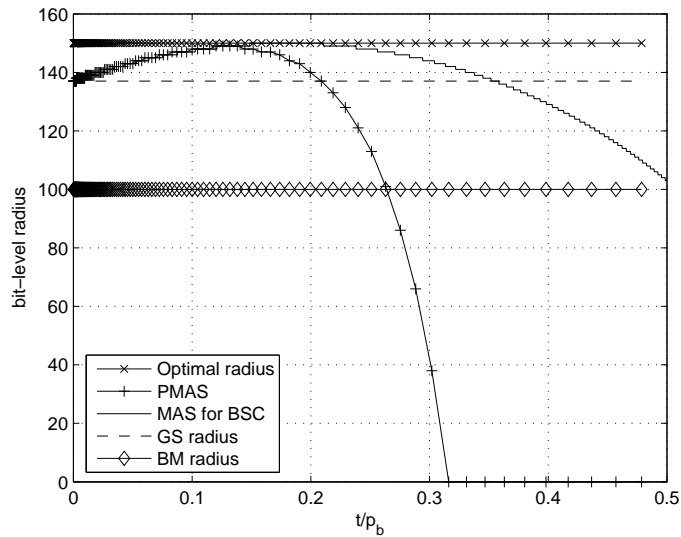}
\caption{Bit-level decoding radius of an RS (255,55) code} \label{fig:radius}
\end{center}
\end{figure}

The error correction radius of optimal MAS as a function of $t$ over an RS(255, 55) code is given in Figure
\ref{fig:radius}. It can be seen that optimal MAS (which is achieved by $t =0.2$) corrects 13 and 50 more
bit-level errors than GS and BM in the worst case. Besides, we also plot bit-level radius of PMAS as a function
of the crossover probability $p_b$ of the BSC. Note that PMAS is not asymptotically optimal for the BSC here.
Even though we choose $p_b$ to maximize the bit-level radius (around $p_b = 0.13$), the bit-level decoding
radius is still 1 bit smaller than that of optimal MAS. The reason can be explained as follows: the worst
case error pattern of this BSC is shown to be all bit-level errors spread in different symbols, thus, the proposed
asymptotically optimal MAS only has to assign multiplicities to candidates of type $0$ and type $1$. On the other
hand, PMAS assigns multiplicities proportionally. Thus it also assigns multiplicities to candidates of all types and unnecessarily spends more cost, which makes it suboptimal in terms of achieving the worst case bit-level decoding radius.

We consider the performance of this asymptotically optimal MAS using a toy example shown in Figure \ref{fig:rs(7,3)}. Consider the performance of an RS(7, 3) code over the BSC. For this code, the decoding radii of conventional BM and GS algorithm are 2 and 3 respectively. Note that, the bit-level decoding radius of PMAS can also be optimized over the crossover probability. However, in this case, the optimal bit-level radius of PMAS is still 3, while the proposed optimal MAS, on the other hand, can achieve bit-level decoding radius 4. We can see from Figure \ref{fig:rs(7,3)}, the performance upper bound of ASD under the proposed optimal MAS outperforms GS and BM.

\begin{figure}[h]
\begin{center}
\includegraphics[width=4.0in]{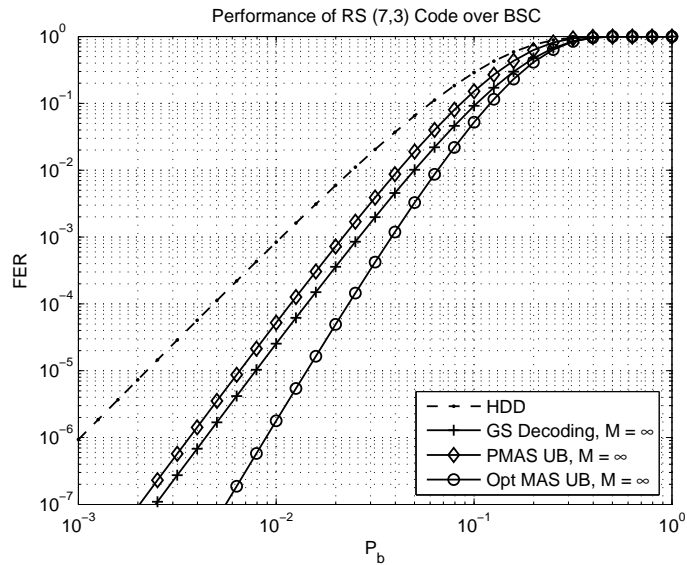}
\caption{Upper Bounds of ASD for RS (7,3) code over the BSC} \label{fig:rs(7,3)}
\end{center}
\end{figure}

\begin{rem}
Over BSC's, the gain of ASD with infinite cost over GS decoding is very little for practical high rate RS codes. Besides, simply increasing the bit-level decoding radius may not necessarily lead to better performance at a moderate FER level. Since GS is a symbol-level decoding algorithm, it may be able to correct more typical error patterns at a moderate FER level than a bit-level decoding algorithm with a slightly larger bit-level decoding radius and hence leads to a better performance at that FER level than ASD with the proposed asymptotically optimal MAS.
\end{rem}

\section{Bit-level Decoding Region of Algebraic Soft-decision Decoding Algorithm}
\label{sec:bit_region}

In this section, we generalize the analysis in previous two sections to a bit-level error and erasure channel. A
simple MAS is proposed and its properties are studied. The decoding region of the proposed MAS in terms of the
number of errors $e$ and the number of bit-level erasures $f$ is investigated for both infinite and finite cost
cases. Finally, we show that the decoding region of the proposed MAS monotonically enlarges as the multiplicity
parameter $M$ increases.

\subsection{The Proposed MAS for a Mixed Error and Erasure Channel} \label{subsec:inf_cost}

We first propose a MAS for the mixed channel, which is motivated by the analysis of the previous two sections.
In Section \ref{sec:rs_erasure}, a simple proposed MAS has been shown to have nearly the same performance as
optimal PMAS for high rate RS codes over the BEC. On the other hand, as shown in Section \ref{sec:rs_bsc}, ASD
even with an optimal MAS has hardly any gain over GS decoder for high rate RS codes over BSC's. Altogether,
these results suggest that most of the gain of ASD for high rate RS codes is from taking care of 1-bit erased
symbols. Therefore, we expect to obtain most of the gain of ASD over other channels comes from assigning nonzero multiplicity only to symbols having zero or one bit erasure. We have the following proposed MAS:

\emph{\underline{Proposed Multiplicity Assignment Strategy}}: In each received symbol, we assign $m_0 = M$ if that symbol does not contain erased bits, assign $m_1 = M/2$ if the symbol contains 1-bit
erasure and assign multiplicity zero for symbols containing more than 1-bit erasures, that is to set $m_j = 0, j
= 2, \cdots, m$.

Under the proposed MAS, there are 5 types of symbols, which are listed below with their corresponding score per
symbol $S_B$ and cost per symbol $C_B$:

(T-1) correctly received symbol: $S_B = M$ and $C_B = \frac{M^2+M}{2}$

(T-2) symbol got erased at symbol-level: $S_B = 0$ and $C_B = 0$

(T-3) erroneous symbol without erasure: $S_B = 0$ and $C_B = \frac{M^2+M}{2}$

(T-4) 1-bit erased symbol without other errors: $S_B = \frac{M}{2}$ and $C_B = \frac{M^2+2M}{4}$

(T-5) 1-bit erased symbol with other errors: $S_B = 0$ and $C_B = \frac{M^2+2M}{4}$

As before, we first characterize the worst case error pattern for the proposed MAS, which dominates the
performance for high rate RS codes. We first have the following lemma:
\begin{lem}\label{lem:f_max_infty}
Under the proposed MAS, an ASD error occurs over the mixed channel if $S \le M(K-1)$.
\end{lem}
\begin{proof} \label{prf:f_max_infty}
When $S \le M(K-1)$, $T(S) \le T(M(K-1))$. Since $T(M(K-1)) = (M+1)M(K-1)/2$ and $T(S)$ is a convex function, it
is easy to verify the following upper bound on $T(S)$:
\begin{equation}
T(S) \le \frac{1}{2}(M+1)S,~~0 \le S \le M(K-1)
\end{equation}
Considering all types of symbols, we have $\frac{1}{2} (M+1) S_B \le C_B$. Therefore, for any received word,
we have the following:
\begin{eqnarray}
T(S) \le \frac{1}{2}(M+1)S \le C
\end{eqnarray}
\end{proof}

Next, we show that recovering bit-level erasures in error-free symbols improves the performance monotonically.
\begin{lem}\label{lem:monotone_erasure_region}
Over the mixed channel, suppose a received vector $\underline{y}$ is certainly decodable by ASD under the proposed MAS with multiplicity parameter $M$. Any other received vector $\underline{y}'$ with only a subset of symbols, which are in error or erased in $\underline{y}$, in error or erased, is certainly decodable under the same condition.
\end{lem}
\begin{proof}\label{prf:monotone_erasure}
Recovering 1 bit-level erasure from an error-free symbol can be of the following 3 cases:

1) From a symbol with more than 2-bit erasures: $\Delta S = 0$ and $\Delta C = 0$

2) From a symbol with exactly 2-bit erasures: $\Delta S = \frac{M}{2}$ and $\Delta C = \frac{M^2+2M}{4}$

3) From a symbol with exactly 1-bit erasure: $\Delta S = \frac{M}{2}$ and $\Delta C = \frac{M^2}{4}$

Obviously, Case 1) is certainly decodable. For Case 2) and Case 3), let $S$, $C$ and $S'$, $C'$ be scores and costs before
and after erasure recovering. We have $a(K-1) \le S \le (a+1)(K-1)$, where $a$ is an integer. Since $T(S) > C$
and according to Lemma~\ref{lem:f_max_infty}, we must have $a+1 > M$. Since $T(S)$ is a piecewise linear
function with monotonically increasing slope, we have:
\begin{eqnarray}
\nonumber   T(S') &=& T\left(S+\frac{M}{2}\right)\\
\nonumber   &\ge& T(S)+(a+1)\frac{M}{2}\\
\nonumber   &>& C(f)+\frac{M^2}{2}\\
\label{eqn:f_decrease_better}   &\ge& C(f)+\frac{M^2+2M}{4} \ge C(f')
\end{eqnarray}
where (\ref{eqn:f_decrease_better}) is due to the fact that $M \ge 2$ in the proposed MAS.
\end{proof}

The following lemma shows that spreading bit-level erasures in different error-free symbols results in a worse
performance than putting them in the same symbol.

\begin{lem}\label{lem:worst_case_fin_erasure}
Over the mixed channel, suppose a received vector $\underline{y}$ has a symbol containing more than 1 bit-level erasures and we
move 1 bit-level erasure from this symbol to a correctly received symbol to get another received vector $\underline{y}'$. If the resulting vector $\underline{y}'$ is certainly decodable using the proposed MAS with multiplicity parameter $M$, then the original received vector $\underline{y}$ is also certainly decodable under the same
condition.
\end{lem}
\begin{proof}\label{prf:worst_case_fin_erasure}
Moving 1 bit-level erasure from a symbol with more than 1-bit erasure to a correctly received symbol can be of
the following 3 cases:

1) From a symbol with more than 2 erasures, $\Delta S = -\frac{M}{2}$ and $\Delta C = -\frac{M^2}{4}$

2) From a symbol with exactly 2 erasures and no errors, $\Delta S = 0$ and $\Delta C = \frac{M}{2}$

3) From a symbol with 2 erasures and some errors, $\Delta S = -\frac{M}{2}$ and $\Delta C = \frac{M}{2}$

Case 1) is nothing but adding 1-bit erasure in a correct symbol. As shown in
Lemma~\ref{lem:monotone_erasure_region}, it results in no better performance. In Case 2) and Case 3), moving
1-bit erasure to correct symbols leads to no larger scores but a larger costs, therefore, it will also result in
no better performance.
\end{proof}

With the above lemmas, we now characterize the worst case error and erasure pattern.
\begin{thm}\label{thm:finite_worst_case0}
Over the mixed channel, the worst case error and erasure pattern for the proposed MAS is that all bit-level errors are spread in different erasure-free symbols and bit-level erasures are spread evenly in the remaining symbols. Besides, if the worst case pattern received vector $\underline{y}$ with $e$ errors and $f$ erasures is certainly decodable under the proposed MAS with multiplicity parameter $M$, any received word $\underline{y}'$ with $e' \le e$ bit-level errors and $f' \le f$ bit-level erasures is certainly decodable under the same condition.
\end{thm}
\begin{proof}
In the worst case, errors should obviously be spread in different symbols. Besides, having erasures in erroneous
symbols will lead to the same score, but a smaller cost. Hence, in the worst case, errors and erasures should
also be spread in different symbols. If the number of errors $e$ and the number of bit-level erasures $f$
satisfy $e+f \le N$, according to Lemma~\ref{lem:worst_case_fin_erasure}, putting erasures in a correctly
received symbol is the worst case. Applying Lemma~\ref{lem:worst_case_fin_erasure} recursively, in the worst
case, bit-level erasures are spread in different symbols. If $e+f > N$, putting more than 2 bit-level erasures
in the same symbol essentially reduces the number of bit-level erasures in error-free symbols and according to
Lemma~\ref{lem:monotone_erasure_region}, it always leads to no worse performance. As a result, when $e+f
> N$, in the worst case, we must have errors, 1-bit erased symbols and 2-bit erased symbols occupying
all $N$ symbols of the received word.

On the other hand, fewer errors will lead to better performance in the worst case. Erasures will only
participate in error-free symbols in the worst case. According to Lemma \ref{lem:monotone_erasure_region}, fewer
bit-level erasures in error-free symbols leads to no worse performance. In conclusion, for any received
word, the worst case for the proposed MAS is that all errors are spread in different erasure-free symbols and erasures are spread evenly in the remaining symbols. Besides, reducing the number of bit-level errors $e$ or the number of bit-level
erasures $f$ will not degrade the worst case performance.
\end{proof}

Theorem~\ref{thm:finite_worst_case0} characterizes the worst case error and erasure pattern, which makes the
decoding region analysis easier.

\begin{cor}
\label{cor:worst_score_cost}
Over the mixed channel, the score and the cost of the proposed MAS with multiplicity parameter $M$ in the worst case can be expressed in terms of the number of errors $e$ and the number of bit-level erasures $f$ as follows:
\begin{align}
S &= \left(N-e-f/2\right)M\\
C &\le (2N-f)\frac{M^2}{4}+N\frac{M}{2}
\end{align}
\end{cor}
\begin{proof}
The corollary is immediate by considering the worst case error and erasure pattern in
Theorem~\ref{thm:finite_worst_case0} in both $e+f \le N$ and $e+f > N$ cases.
\end{proof}

\begin{cor}
\label{cor:worst_e_f_region}
Over the mixed channel, an ASD error occurs under the proposed MAS if $f~\ge~2(N-(K-1)-e)$
\end{cor}
\begin{proof}
The corollary is obtained by combining Lemma~\ref{lem:f_max_infty} and Corollary~\ref{cor:worst_score_cost}.
\end{proof}
Corollary~\ref{cor:worst_e_f_region} suggests that an ASD error occurs under the proposed MAS before all error-free
symbols get erased at the symbol-level. Besides, it also gives an outer bound on the decoding region of the
proposed MAS. The exact decoding region of the proposed MAS will be studied in more detail in the following
subsection.

\subsection{Infinite Cost Performance Analysis} \label{subsec:inf_cost}

Due to the simplicity, the decoding region of this proposed MAS for medium to high rate RS codes can be
characterized analytically. First, we consider the infinite cost case.

\begin{thm}\label{thm:bgmd_mas}
Under the proposed MAS with $M \rightarrow \infty$, the decoding region over the mixed channel in terms of $e$
and $f$ when $e+f \le N$ is:
\begin{equation}
e < N-f/2-\sqrt{(K-1)(N-f/2)}
\end{equation}
\end{thm}
\begin{proof} \label{prf:bgmd_mas}
When $e+f \le N$, in the worst case the score and the cost can be expressed as
\begin{eqnarray}
\label{eqn:inf_score} S &=& (N-e-f/2)M\\
\label{eqn:inf_cost} C &=& 1/4 M^2(1+o(1)) (2N-f)
\end{eqnarray}
Plugging in (\ref{eqn:sufficient}), we can get:

\begin{equation}
\label{eqn:inf_region} e < N-f/2-\sqrt{(K-1)(N-f/2)}
\end{equation}
\end{proof}
According to Corollary~\ref{cor:worst_score_cost}, when $e+f > N$, (\ref{eqn:inf_region}) is still achievable
and the actual decoding region can be larger. When $f = 0$, the above region becomes the maximum error
correcting radius of GS decoding; when $e = 0$, we can obtain the worst case bit-level decoding radius derived
in (\ref{align:opt_e_a}).

To get an idea on how good this proposed MAS is, we derive an outer bound on the optimal decoding region of ASD
with infinite cost. Using a technique similar to that used in Section \ref{sec:rs_bsc}, we first derive an optimal MAS over a 1-bit flipped or erased channel. That is, we assume in each symbol of the RS codeword, there is at most either 1 bit in error or at most 1-bit erasure. The derivation of the optimal decoding region for this channel is given in Appendix \ref{pdx:1bit_region}. In general, the 1-bit flipped or erased channel is optimistic compared with the actual bit-level error and erasure channel. Hence, when $e+f \le N$, the optimal
decoding region of a 1-bit flipped or erased channel serves as an outer bound of the actual decoding region of a
mixed error and bit-level erasure channel.

\subsection{Finite Cost Performance Analysis} \label{subsec:fin_cost}

Consider, the proposed MAS with finite cost, in the simplest case, $M = 2$. That is, we assign $m_0 = 2$ to
symbols without erasures; if there is 1 bit-level erasure, we assign $m_1 = 1$ to each candidate;
otherwise, we assign $m_i = 0, i = 2, 3, \cdots, m$. The decoding region is characterized in the following
theorem.

\begin{thm}\label{thm:bgmd_finite}
Under the proposed MAS with $M = 2$, the decoding region of RS codes of rate $R \ge 2/3+1/N$ over the mixed
channel is:
\begin{equation}
\label{eqn:fin_region} e < \frac{1}{2}(N-K+1)-\frac{f}{3}
\end{equation}
\end{thm}
\begin{proof} \label{prf:bgmd_finite}
For $R \ge 2/3+1/N$, in the worst case, errors and erasures will not overlap. Hence, $S = 2(N-e-f/2)$ and $C =
3N-f$. We must have:
\begin{align}
&(a+1) (2(N-e-f/2)-a/2(K-1))> 3N-f\\
&a(K-1) < 2(N-e-f/2) \le (a+1)(K-1), \text{~where~a is a non-negative integer}
\end{align}

For $a = 0, 1$, we get contradictions.

For $a \ge 3$, we get trivial bounds.

For $a = 2$, we obtain the decoding region:
\begin{equation}
\label{eqn:bgmd_finite} e < \frac{1}{2}(N-K+1)-\frac{1}{3} f, \text{~for~} (K-1)/N \ge 2/3
\end{equation}
\end{proof}

\begin{cor}
\label{cor:high_rate_approx} For RS codes of rate $R < 2/3+1/N$, the decoding region over the mixed channel in
Theorem~\ref{thm:bgmd_finite} is achievable under the proposed MAS.
\end{cor}
\begin{proof}
Since $T(S)$ has a monotonically increasing slope, when (\ref{eqn:fin_region}) is satisfied, we must have $T(S)
> C$ if $e+f \le N$. If $e+f > N$, again due to Corollary~\ref{cor:worst_score_cost}, the above region is still
achievable.
\end{proof}

We can also derive a decoding region of the proposed MAS with any given multiplicity parameter $M$ as follows:
\begin{thm}\label{thm:finite_region}
Under the proposed MAS with a given multiplicity parameter $M$, the decoding region in terms of the number of errors $e$
and the number of bit-level erasures $f$ (where $e+f \le N$) can be expressed as:

\begin{equation}
\label{eqn:bgmd_fin_M} e < N-\frac{f}{2}-\frac{\hat{a}(\hat{a}+1)(K-1)/2+C}{M(\hat{a}+1)}
\end{equation}
with the cost $C = \frac{1}{2}(N-f)M(M+1) + f \frac{M}{2}(\frac{M}{2}+1)$ and $\hat{a} = \lfloor
\frac{-1+\sqrt{1+\frac{8C}{K-1}}}{2}\rfloor$.
\end{thm}
The derivation of Theorem \ref{thm:finite_region} is provided in Appendix \ref{pdx:finite_region}. Similarly
when $e+f >N$, the region in Theorem~\ref{thm:finite_region} is still achievable. Though the actual decoding
region can be even larger for low rate RS codes, we are not going study it in more detail in this paper.

We give some examples of the decoding region of RS codes over the mixed channel under different decoding
schemes. In Figure \ref{fig:region_rs(255,239)}, a high rate RS(255, 239) code is considered. The decoding region of BM and GS decoding are exactly the same. However, when considering erasures at the bit-level, the decoding region of the proposed MAS is significantly larger than BM and GS decoding. For the proposed MAS with $M = 2$, the simulated bit-level region (obtained by numerically checking the sufficient condition for each error and erasure pair), the high rate bit-level achievable region as in Theorem~\ref{thm:bgmd_finite} and the general achievable region as in Theorem~\ref{thm:finite_region} are right on top of each other. The maximum number of bit-level erasures the proposed MAS with $M = 2$ can correct is about 1.5 times that of BM and GS decoding at the bit-level and the decoding region of the proposed MAS with $M = 2$ is significantly larger than BM and GS decoding region. In the infinite cost case, the proposed MAS achieves the outer bound of the decoding region. The decoding region corresponding to $M = \infty$ encompasses the decoding region corresponding to $M = 2$.

\begin{figure}[h]
\begin{center}
\includegraphics[width=4.0in]{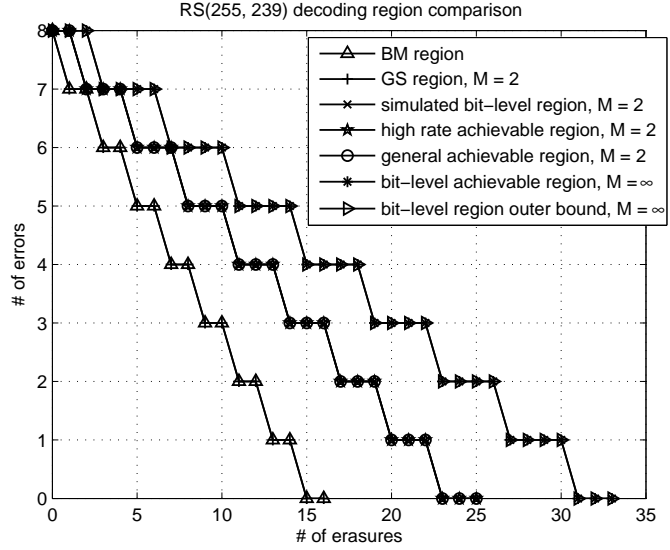}
\caption{Bit-level Decoding Region of Algebraic Soft Decoding for RS(255, 239)} \label{fig:region_rs(255,239)}
\end{center}
\end{figure}

In Figure \ref{fig:region_rs(63,23)}, we show the decoding region of a low rate code RS($63, 23$). In this case,
the high rate achievable region in Theorem~\ref{thm:bgmd_finite} becomes loose. On the other hand, the
general decoding region derived in Theorem \ref{thm:finite_region} still coincides with the actual decoding
region (by checking the sufficient condition for each error and erasure pair). When there is no erasure, the
maximum number of errors the proposed MAS can correct is the same as GS decoding. Again, since ASD can take
advantage of the erasure information at the bit-level, the decoding region of the proposed MAS is strictly
larger than the decoding region of GS with symbol-level error and erasure decoding. When $e+f > N$ in the
infinite cost case, the outer bound becomes invalid. However, the achievable region in the infinite cost case is
still a valid achievable region.

\begin{figure}[h]
\begin{center}
\includegraphics[width=4.0in]{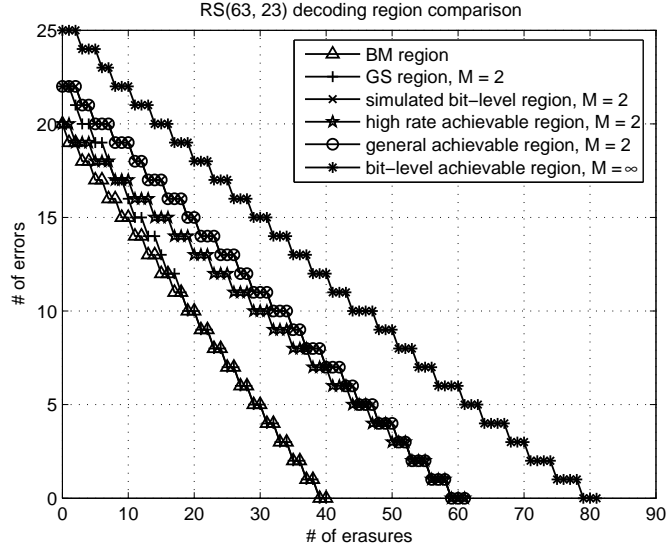}
\caption{Bit-level Decoding Region of Algebraic Soft Decoding for RS(63, 23)} \label{fig:region_rs(63,23)}
\end{center}
\end{figure}

\subsection{Monotonicity} \label{subsec:monotonicity}

In this subsection, we show the monotonicity of the decoding region of the proposed MAS as a function of
multiplicity parameter $M$ over the mixed channel. It was shown by McEliece in \cite{mceliece_gsd}, the error correction
radius of GS algorithm is a monotonic function of multiplicity parameter $M$. This monotonicity does not hold for ASD
algorithms in general. However, the monotonicity result is of interest since it justifies that the asymptotical
performance analysis by letting $M \rightarrow \infty$ is indeed the ``best'' achievable result and it also
verifies that increasing the cost will lead to at least no worse performance.

We need the following property of the function $T(S)$:

\begin{lem}\label{lem:T(s)}
$T((a+1)x) \ge \frac{a+2}{a} T(x)$, if $x \ge K-1$ and $a$ is a positive integer.
\end{lem}
\begin{proof} \label{prf:T(s)}
This proof of this lemma is similar to Theorem A-1, (A-9) in \cite{mceliece_gsd}.
Since $x \ge K-1$, we have:
\begin{align}
\left(1+\frac{l}{a}\right)(K-1) \le &x \le \left(1+\frac{l+1}{a}\right)(K-1) \text{~~for~~} l = 0, 1, 2, \cdots, a-1\\
\label{eqn:ax} (a+l)(K-1) \le &a x \le (a+l+1)(K-1)
\end{align}
$T(a x)$ can be computed as:
\begin{equation}
\label{eqn:T(ax)}
 T(a x) = (a+l+1)\left[a x-\frac{a+l}{2}(K-1)\right]
\end{equation}
On the other hand, $(a+1)x$ is in the following range as:
\begin{eqnarray}
\frac{a^2+(l+1)a+l}{a}(K-1) &\le (a+1)x \le& \frac{a^2+(l+2)a+(l+1)}{a}(K-1)\\
\label{eqn:(a+1)x}(a+l+1)(K-1) &\le (a+1)x \le& (a+l+3)(K-1)
\end{eqnarray}
Since $T(S)$ is a piecewise linear function with monotonically increasing slope, $T(S) \ge
(i+1)(S-\frac{i}{2}(K-1))$ for any non-negative integer $i$. Hence, we have the following lower bound on
$T((a+1)x)$:
\begin{equation}
\label{eqn:T((a+1)x)}T((a+1)x) \ge (a+l+2)\left[(a+1)x-\frac{a+l+1}{2}(K-1)\right]
\end{equation}

Combining (\ref{eqn:T(ax)}) and (\ref{eqn:T((a+1)x)}), we have the following:
\begin{eqnarray}
\nonumber T((a+1)x)-\frac{a+2}{a}T(a x) &\ge& (a+l+2)\left[(a+1) x-\frac{a+l+1}{2}(K-1)\right]\\
&&-\frac{a+2}{a}(a+l+1)\left[a x-\frac{a+l}{2}(K-1)\right]\\
&\ge& -l x+\frac{(a+l+1)l}{a}(K-1)\\
\label{eqn:T_gap}&=& \frac{l}{a}\left[(K-1)(a+l+1)-ax\right] \ge 0
\end{eqnarray}
where the final step in (\ref{eqn:T_gap}) follows by the fact that $l \ge 0$ and $ax \le (K-1)(a+l+1)$.
\end{proof}

\begin{thm}\label{thm:monotonicity}
Over the mixed channel, if a received word is certainly decodable using ASD with multiplicity parameter $M$, it is certainly decodable under multiplicity parameter $M+2$ ($M$ has to be even in the proposed MAS), which means the performance of ASD under the proposed MAS is monotonic with the multiplicity parameter $M$.
\end{thm}
\begin{proof} \label{prf:monotonicity}
If a codeword is certainly decodable with the multiplicity parameter $M$, we have $T(S(M)) > C(M)$, where $S(M)$ and $C(M)$ are score and cost with multiplicity $M$ respectively. Considering all types of symbols in the received word, we have
the following relationship:
\begin{eqnarray}
   S(M+2) &=& \frac{M+2}{M}S(M)\\
\label{eqn:cost_relation}C(M+2) &\le& \frac{(M+2)(M+3)}{M(M+1)}C(M)
\end{eqnarray}

If a received word is certainly decodable, according to Lemma~\ref{lem:f_max_infty}, we have $S(M) > M(K-1)$.
Therefore:
\begin{eqnarray}
\nonumber  T(S(M+2)) &=& T((M+2)\frac{S(M)}{M})\\
\label{eqn:sm1}  &\ge&   \frac{(M+3)(M+2)}{(M+1)M}T(S(M))\\
\nonumber       &>&   \frac{(M+3)(M+2)}{(M+1)M}C(M)\\
\label{eqn:sm2} &\ge&   C(M+2)
\end{eqnarray}
where (\ref{eqn:sm1}) is obtained by applying Lemma~\ref{lem:T(s)} twice and (\ref{eqn:sm2}) is due to
(\ref{eqn:cost_relation}).
\end{proof}
Note that the monotonicity property holds for all RS codes regardless of the rate.

\section{Bit-Level Generalized Minimum Distance Decoding Algorithm}
\label{sec:bit_gmd}

In this section, we develop a practical SDD algorithm for RS codes, which is motivated by the analytical results
in the previous sections.

\subsection{The Generic BGMD Algorithm}
\label{subsec:bgmd_algorithm}

As shown in Section~\ref{subsec:fin_cost}, the proposed MAS has a significantly larger decoding region than
conventional BM and GS decoding over a mixed error and bit-level erasure channel. This provides the intuition
that properly treating erasures at the bit-level will also help in RS soft-decision decoding over other
channels. An efficient way to utilize erasures over many channels is by ordering the reliability values of the
received bits, treating the LRB's as erasures and running an error and erasure decoder successively, namely
generalized minimum distance (GMD) decoding \cite{forney_gmd}. In each iteration, the decoder can decode
erasures in the LRB's together with some extra errors in the remaining most reliable bits (MRB's) as long as the
error and erasure $\text(e, f)$ pair is within the decoding region of BM algorithm. Due to the similarity
between the proposed algorithm and conventional symbol-level GMD for RS codes, it is called bit-level GMD
(BGMD).

The generic algorithm of BGMD is described in Algorithm \ref{alg:bgmd}.

\begin{algorithm}[h]
\begin{description}
\item[\textbf{Step1.}] Initialization: set the initial iteration round $i = 1$ and generate the log likelihood
ratio (LLR) for each coded bit based on the channel observation $y_j$: $L_j = \log{\frac{P(c_j = 0|y_j)}{P(c_j =
1|y_j)}}$, for $j = 1, 2, \cdots, n$

\item[\textbf{Step2.}] Reliability Ordering: order the coded bits according to the absolute value of the LLR's
$\{|\emph{L}_j|\}$ in ascending order and record the ordering indices $\{s_j\}$.

\item[\textbf{Step3.}] Bit-level Hard Decision:
        $\hat{c_j}=\left\{%
        \begin{array}{ll}
            0, & \hbox{$L_j > 0$;}\\
            1, & \hbox{$L_j \le 0$.}\\
        \end{array}%
        \right.$

\item[\textbf{Step4.}] Multiplicity Assignment:

        In each symbol of the received word $\underline{\hat{c}}$, assign multiplicities according to:

        1) if no bit is erased, assign $M$ to the received symbol;

        2) if there is 1-bit erasure, assign $M/2$ to each candidate;

        3) if there is more than 1-bit erasure, assign multiplicity zero.

\item[\textbf{Step5.}] Algebraic Soft Decision Decoding: Run ASD according to the multiplicity assignment
determined in \textbf{Step4.}. Keep the generated codewords in the decoding list.

\item[\textbf{Step6.}] Erase the Least Reliable $i$ Bits: $\hat{c}_{s_{l}} = \epsilon$ for $l = 1, \cdots, i$, (where $\epsilon$ indicates a bit-level erasure).

\item[\textbf{Step7.}] Iteration:
        If $i \le n-k$ and ASD is still able to correct the current erasures given no error, set $i \leftarrow i+1$ and go to \textbf{Step4.} for another decoding iteration.

\item[\textbf{Step8.}] Final Decision:
        Output the most likely codeword in the decoding list. If there is no codeword in the list, a decoding failure is declared.
\end{description}
\caption{Bit-level Generalized Minimum Distance Decoding Based on Algebraic Soft Decision Decoding for
Reed-Solomon Codes} \label{alg:bgmd}
\end{algorithm}

\begin{rem}
In terms of implementation, BGMD does not need to run ASD algorithm many times. In fact, the interpolation part
can be shared between different erasure patterns. Similar to the techniques proposed in \cite{ahmed_isit,
haitao_isita, haitao_thesis, bellorado_kvchase}, we can generate all the candidate codewords in one
interpolation round by applying factorization in the intermediate steps during the interpolation procedure.
Besides, factorization needs to be performed only at outer corner (e, f) points. For high rate RS codes, the
number of ``test erasure patterns'' of BGMD is the same as conventional symbol-level GMD.
\end{rem}

\subsection{Performance Analysis of BGMD}
\label{subsec:bgmd_bound}

Due to the simple structure of BGMD, the performance of BGMD for practical high rate RS codes over an AWGN channel can be tightly bounded using order statistics techniques. Define
$D(M)$ as the decoding region of the proposed MAS over a mixed bit-level error and erasure channel,
namely the set of error and erasure $(e, f)$ pairs that is certainly decodable by the proposed MAS with multiplicity parameter $M$ as specified in Theorem~\ref{thm:finite_region}. Let $f_{max, M}$ and $e_{max, M}$ be the maximum number of errors and erasures respectively such that $(0, f_{max, M})$ and $(e_{max, M}, 0)$ are still in $D(M)$. The FER
of BGMD can be upper bounded by the FER performance of using a set of bit-level error and erasure decoders, each
with different number of erased bits $f$ in the LRB's and a different error correction capability $e$ such that $(e, f) \in D(M)$. Note, however, $D(M)$ is the worst case decoding
region of the proposed MAS, BGMD can in fact correct even more number of errors and erasures if some of the
errors and erasures overlap in some symbols. However, for high rate RS codes, this upper bound becomes tight,
since the worst case error and erasure pattern dominates. Performance analysis of BGMD then boils down to
bounding the performance of a conventional GMD decoder for binary codes \cite{agrawal_gmd} with a skewed
decoding region $D(M)$. Hence, upper bounds of GMD for binary codes, such as the one derived in
\cite{agrawal_gmd}, are directly applicable to evaluating the performance of BGMD decoding. For readers'
convenience, we give the detailed procedure to compute the FER upper bound on BGMD algorithm in Appendix
\ref{pdx:bgmd_bound}. For more comprehensive studies on this bound, we refer interested readers to
\cite{agrawal_gmd} and \cite{fossorier_bound} for applications to other order statistics based decoding
algorithms.

\begin{figure}[h]
\begin{center}
\includegraphics[width=4.0in]{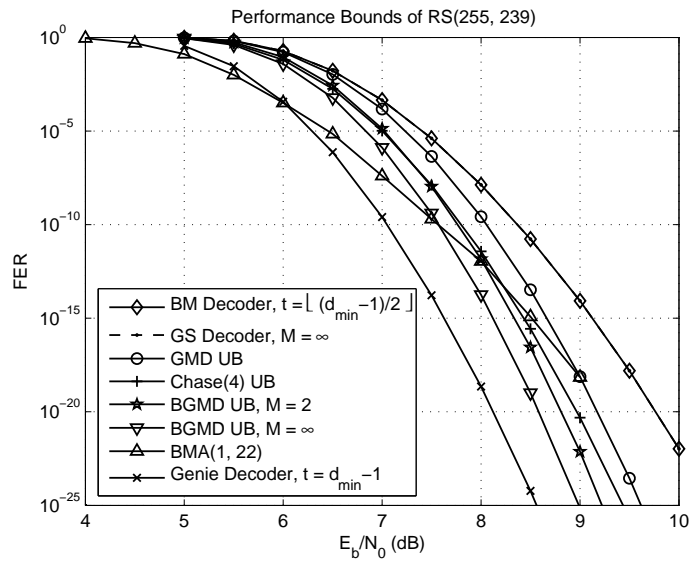}
\caption{Performance Bounds of Algebraic Soft-decision Decoding for RS Code (255, 239) over an AWGN Channel}
\label{fig:bgmd_rs239}
\end{center}
\end{figure}

Thanks to this upper bound, the performance of BGMD in high SNR's, where RS codes operate in many practical
systems, can be predicted analytically, which is beyond the capability of computer simulation. As an example,
performance bound of BGMD over a popular high rate RS(255, 239) is plotted in Figure \ref{fig:bgmd_rs239}. At an
FER = $10^{-14}$, the upper bound of BGMD with $M = 2$ has a 0.8dB and 0.3dB gain over conventional BM and GMD
upper bound respectively. With asymptotically large cost $M = \infty$, the gain of BGMD upper bound over BM
increases to 1dB at an FER = $10^{-14}$. On the other hand, the performance of KV algorithm cannot be simulated
at such a low FER. Compared with another popular SDD algorithm, i.e., the box and match algorithm (BMA) order-1
with 22 bits in the control band \cite{fossorier_bma}, the upper bound of BGMD with $M = 2$ has a 0.2dB gain at
this FER level with a much smaller complexity and memory consumption than BMA. In high SNR's, the upper bound of
BGMD with $M = 2$ also has comparable performance to the upper bound of Chase Type-2 Decoding \cite{chase_chase}
with 16 test error patterns. The performance gap of BGMD to a genie decoder with decoding radius $t = N-K$
becomes smaller and smaller as SNR increases. Note that the actual performance of BGMD may be even better than
that predicted by the upper bound as will be shown in the simulation results in the following section. The FER
upper bound on BGMD can be further tightened by considering the joint order statistics, which also increases
computational complexity.

The generic BGMD algorithm can also be extended to incorporate Chase type decoding \cite{chase_chase,
bellorado_kvchase, haitao_thesis}. Under the proposed MAS, the corresponding performance can also be tightly
upper bounded by similar bounding techniques using order statistics as shown in \cite{fossorier_bound}.

\subsection{Discussions}
\label{subsec:discussion}

We first discuss a counter-intuitive phenomenon of KV decoding, which was first observed in \cite{assaf_thesis}. That is, KV decoder may fail even when the received vector does not contain any errors. We give an example as follows:

\underline{Example~1}: Consider an RS(255, 239). Suppose in the received vector, no bit is in error. $255\times
7 = 1785$ bits are perfectly received, i.e., the magnitude of their LLRs are all infinity. In each symbol, there
is one bit that is corrupted by some noise, but still they have the correct signs. The APP that it is the
transmitted bit is $0.7$ and the probability that it is the wrong bit is $0.3$. According to the KV algorithm,
i.e., PMAS, we will have $S = 178.5 M$ and $C \approx 73.95 M^2$. It is easy to verify that even though there is
no bit in error, the sufficient condition (\ref{eqn:sufficient}) will be violated. It can also be verified by actual simulation that KV will fail in some cases even when no bit is in error. In fact, this phenomenon was
recently reassured in \cite{duggan_asd}. The analysis in \cite{duggan_asd} showed that under PMAS, the
asymptotical decoding radius of ASD might be 0, which suggests the decoder can fail even though there is no error.

At first glance, this phenomenon seems counter-intuitive. It seems to suggest that soft information even
degrades the performance. However, from the analysis in previous sections, we can get an intuitive and sensible
interpretation. ASD in some sense treats weighted erasures, therefore, similar to erasure decoding over AWGN
channels, in some cases, we may end up erasing too many correct bits and cause a decoding failure even though there is no error. On the other hand, since BGMD treats erasures according to the received reliability value and also erases bits successively, these abnormal cases will be excluded.

Besides, in general, the monotonicity of ASD is not guaranteed. For instance, it is observed in \cite{gross_kv}
that for the simplified KV algorithm, the decoding performance does not monotonically improve as the cost
increases. For the proposed BGMD, on the other hand, as shown in the previous section, the decoding region will
monotonically become larger as a function of the multiplicity parameter $M$.

The generic BGMD can naturally be generalized to take more than 1-bit erasures into account, which will be
important in decoding medium to low rate RS codes. The associated performance bounds are also of great research
interest, since for medium to low rate RS codes, the upper bound considering the worst case bit-level decoding
region alone becomes loose.

\section{Simulation Results}
\label{sec:simulation}

In this section, we show simulation results of the proposed BGMD over various communication channels. We will
see that the proposed BGMD, though derived from a simple MAS, is superior to many existing MAS's which are far
more complicated. Besides, in contrast to most MAS's in the literature, the order statistics based upper bound
can accurately evaluate the actual performance of BGMD for many practical high rate RS codes. We assume that
binary phase shift keying (BPSK) is used as the modulation format in all the simulations in this paper, since in this case bit-level soft information can be generated straightforwardly. The proposed BGMD algorithm can also be applied to higher order modulations to provide extra gain over symbol-level soft decision decoding. However, this is beyond the scope of this paper and will be discussed in more detail elsewhere.

\begin{figure}[h]
\begin{center}
\includegraphics[width=4.0in]{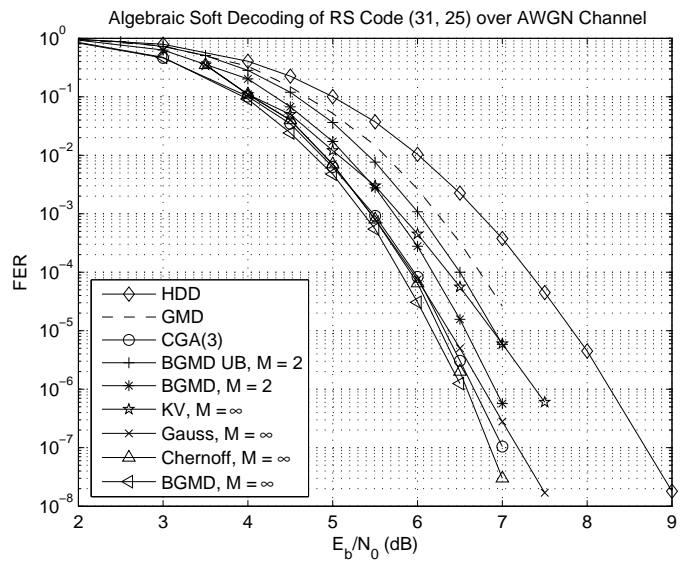}
\caption{Algebraic Soft-decision Decoding of RS Code (31, 25) over an AWGN Channel} \label{fig:rs3125_awgn}
\end{center}
\end{figure}

In Figure \ref{fig:rs3125_awgn}, we plot the FER performance of an RS(31, 25) over an AWGN channel.
BGMD ($M = 2$) outperforms conventional BM by 1.3dB at an FER = $10^{-6}$. It also outperforms conventional symbol-level
GMD by $0.6$dB at an FER = $10^{-5}$ and is slightly inferior to Combined Chase and GMD CGA(3), which has a much
larger complexity, by 0.2dB. Compared with existing MAS's for ASD, it gives favorable performance as well. With
$M = 2$, it even outperforms KV algorithm with $M = \infty$ by 0.5dB at an FER $= 10^{-6}$. With $M = \infty$,
the performance of BGMD outperforms the performance of Gaussian approximation based MAS \cite{parv_gauss} and
the performance of Chernoff technique based MAS \cite{el-khamy_mas, el-khamy_kv}, which are far more complicated
than the proposed BGMD in multiplicity assignment.

\begin{figure}[h]
\begin{center}
\includegraphics[width=4.0in]{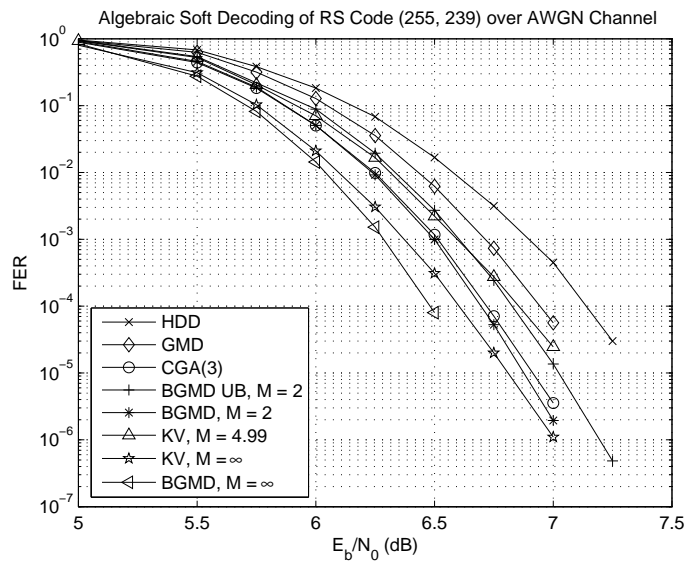}
\caption{Algebraic Soft-decision Decoding of RS Code (255, 239) over an AWGN Channel} \label{fig:rs239_awgn}
\end{center}
\end{figure}

In Fig.~\ref{fig:rs239_awgn} we evaluate the FER performance of a long code, RS(255,239) code. Again, BGMD ($M = 2$) outperforms GMD and is comparable to CGA(3). As the codeword length increases, KV algorithm becomes
asymptotically optimal as shown in \cite{koetter_kv}. The performance of the proposed BGMD is still comparable
to KV decoding. In the infinite cost case, the performance of BGMD ($M = \infty$) is slightly better than the
performance of KV ($M = \infty$); in the finite cost case, BGMD ($M = 2$) even outperforms KV ($M = 4.99$).
Besides, since BGMD only assigns multiplicities to symbols with at most 1-bit erasure, the memory consumption in storing the assigned multiplicities is much smaller than KV. The upper bound is quite tight and it starts to outperform KV ($M = 4.99$) and is only $0.1$dB inferior to the actual performance at an FER = $10^{-5}$. As shown in Figure \ref{fig:bgmd_rs239}, it gives an estimate of the performance of BGMD in high SNR's as well.

\begin{figure}[h]
\begin{center}
\includegraphics[width=4.0in]{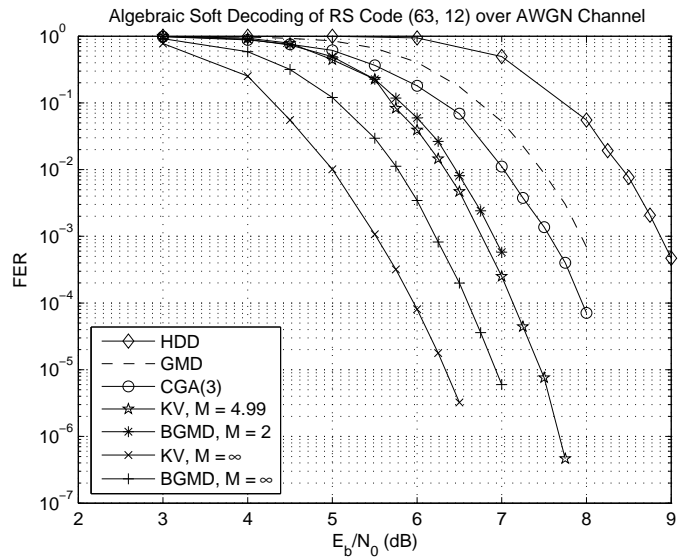}
\caption{Algebraic Soft Decoding of RS Code (63, 12) over an AWGN Channel} \label{fig:rs6312awgn}
\end{center}
\end{figure}

Though the upper bound of BGMD is tight only for medium to high rate RS codes, the proposed BGMD algorithm actually provides even more significant coding gain for low rate RS codes. As shown in Figure \ref{fig:rs6312awgn}, the performance of BGMD ($M = 2$) can outperform BM, GMD, CGA(3) decoding by a large margin for an RS(63, 12) code over an AWGN channel. The gain of BGMD over BM is about 2dB at an FER $= 10^{-4}$. In this case, CGA(3) is far more inferior to BGMD. BGMD ($M = 2$) has almost identical performance as KV ($M = 4.99$). While, in the infinite cost case, KV does have a 0.6dB gain over BGMD at an FER = $10^{-5}$, which suggests that taking care of more than 1-bit-erased symbols might provide extra gains for low rate RS codes. It is an interesting open problem to develop such kind of MAS.

\begin{figure}[h]
\begin{center}
\includegraphics[width=4.0in]{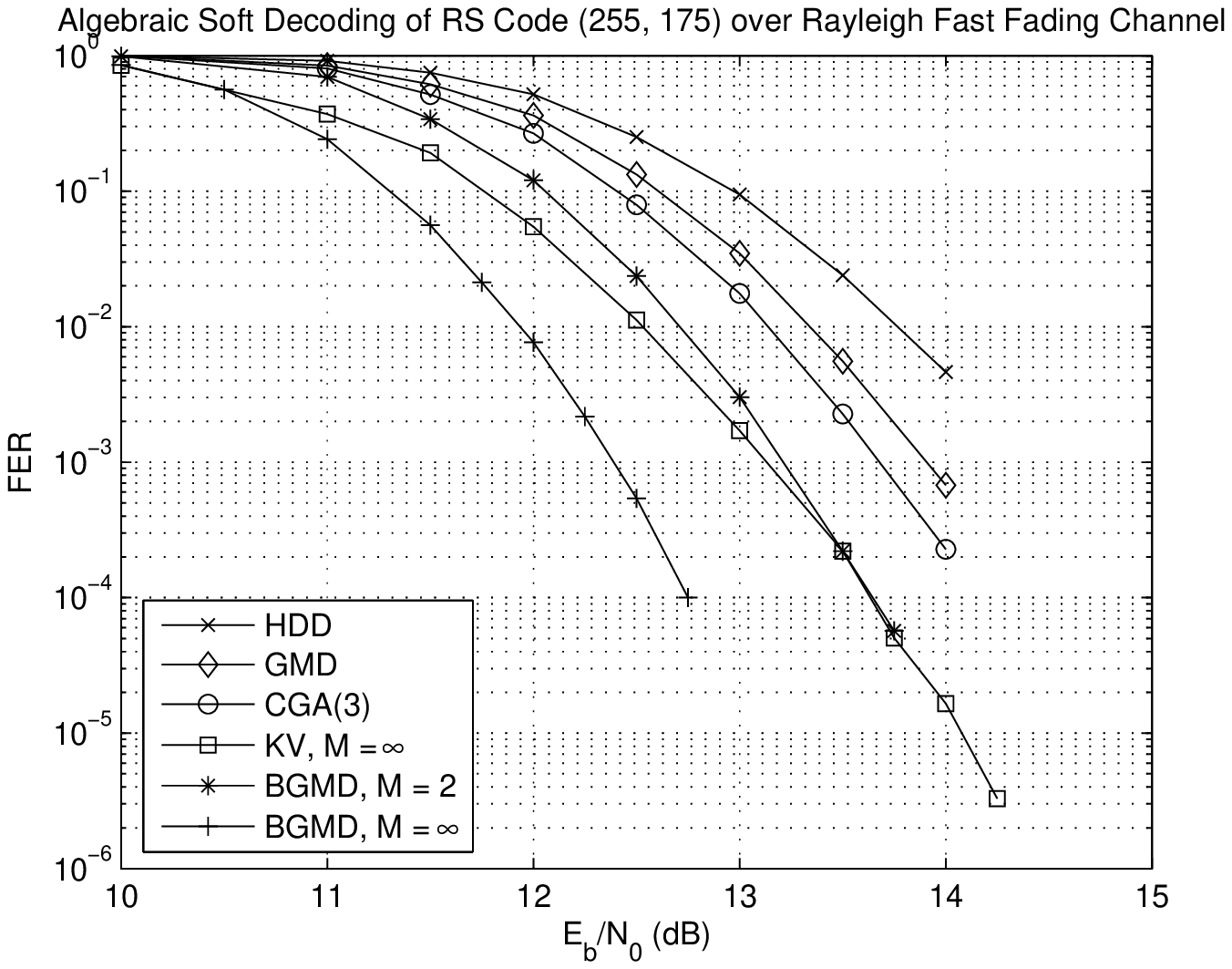}
\caption{Algebraic Soft Decoding of RS Code (255, 175) over the Rayleigh Fast Fading Channel}
\label{fig:rs175_fad}
\end{center}
\end{figure}

The gain of the proposed BGMD over BM and CGA becomes larger when the channel is ``similar'' to a BEC, say Rayleigh
fast fading channels, since BGMD can correct a significantly larger number of bit-level erasures than
conventional BM as discussed in Section \ref{sec:rs_erasure}. As shown in Figure \ref{fig:rs175_fad}, the gain
of BGMD ($M = 2$) is about 1.5dB compared with BM at an FER = $10^{-3}$. As expected, the gain of BGMD over
CGA(3) is more significant over the fading channel. Compared with KV ($M = \infty$), BGMD ($M = 2$) is slightly
inferior to KV ($M = \infty$) in low SNR's, but it intersects KV ($M = \infty$) at an FER = $10^{-4}$ and
performs better in high SNR's. BGMD ($M = \infty$) has a 0.75dB gain over KV ($M = \infty$) at an FER =
$10^{-4}$. The superior performance of BGMD seems to suggest that for high rate RS codes, efficiently taking
advantage of bit-level erasures exploits most of the gain in ASD.


Performance of the proposed BGMD is also investigated over practical magnetic recording channels, that is,
longitudinal channel and perpendicular channel with $90\%$ jitter noise. More details of the channel model can
be found in \cite{kurtas_book}. Similar performance gains of BGMD have also been observed over practical
recording channels. BGMD ($M = 2$) outperforms conventional GMD and performs competitively with KV and CGA(3),
which are much more complex. This superior performance of BGMD suggests that though RS codes are usually
considered as a powerful burst error correction code, it is still beneficial to taking advantage of soft
information at the bit-level even over practical magnetic recording channel models, where errors are usually
bursty.


\begin{figure}[h]
\begin{center}
\includegraphics[width=4.0in]{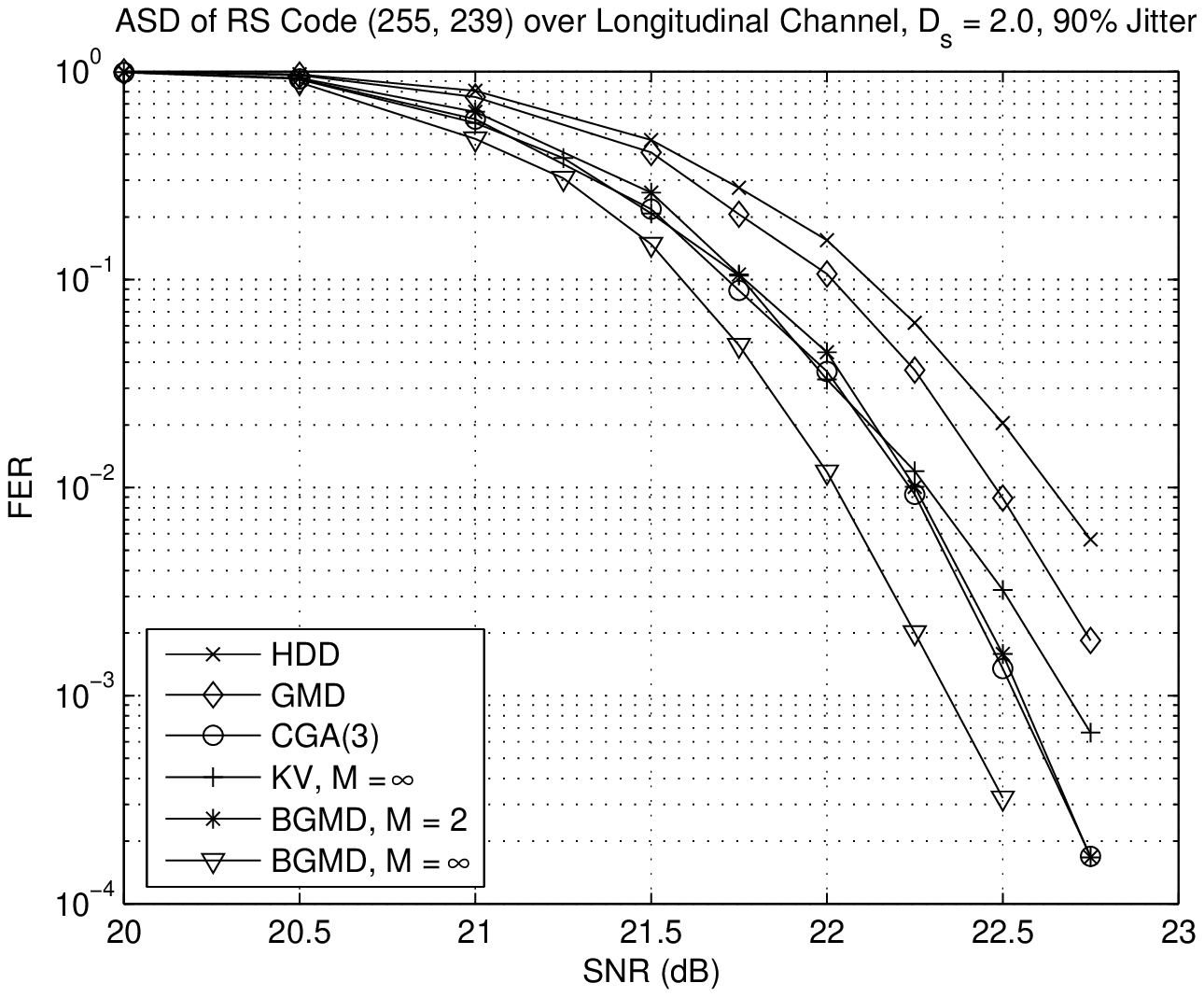}
\caption{Algebraic Soft-decision Decoding of RS Code (255, 239) over Longitudinal Channel, $D_s$ = 2.0, 90$\%$
Jitter} \label{fig:rs239ljitter}
\end{center}
\end{figure}

\begin{figure}[h]
\begin{center}
\includegraphics[width=4.0in]{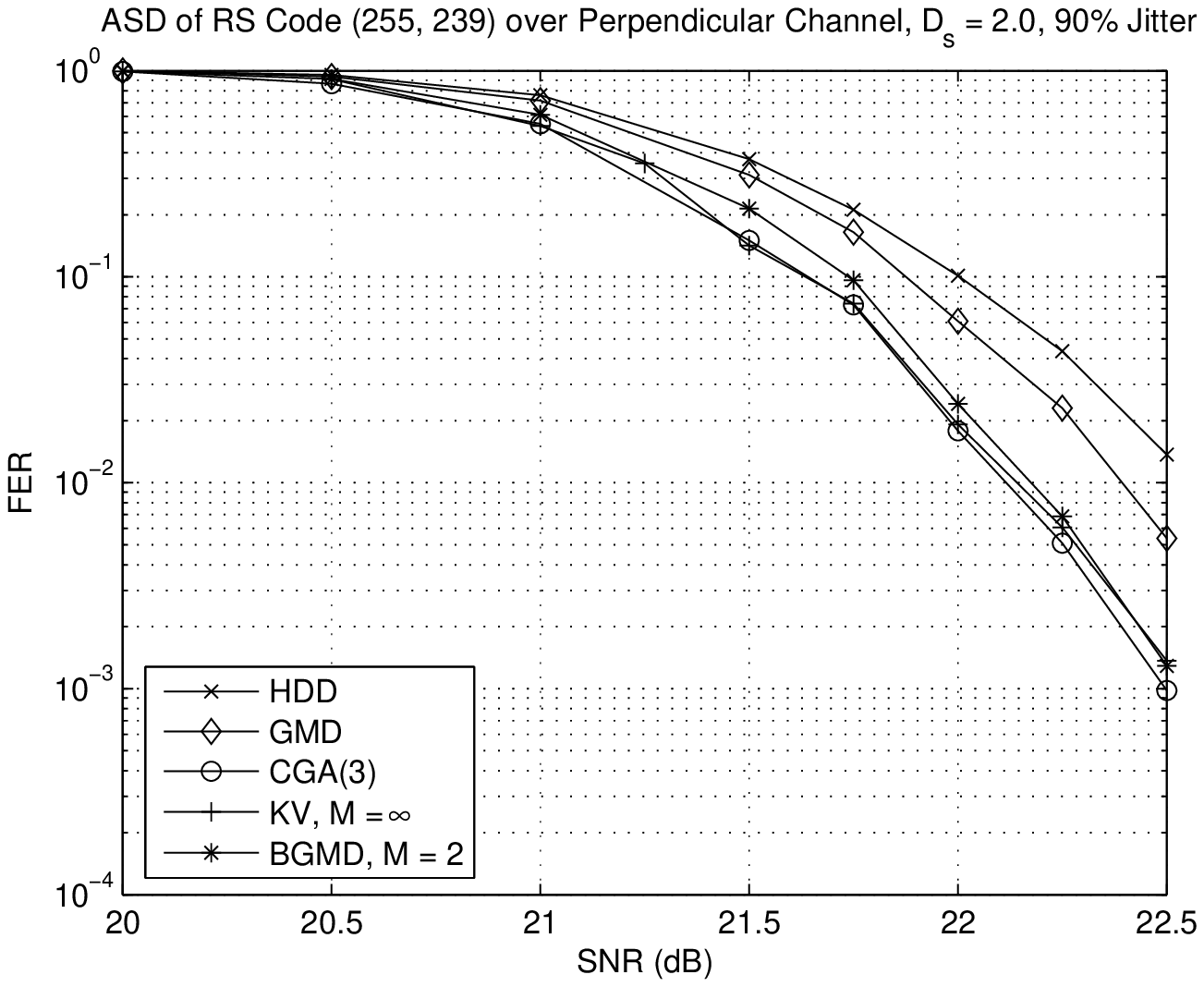}
\caption{Algebraic Soft-decision Decoding of RS Code (255, 239) over Perpendicular Channel, $D_s$ = 2.0, 90$\%$
Jitter} \label{fig:rs239p}
\end{center}
\end{figure}

\section{Conclusion}
\label{sec:conclusion}

We have presented multiplicity assignment strategies and performance analyses of algebraic soft-decision
decoding over erasure channels, binary symmetric channels and mixed error and bit-level erasure channels.
Performance analysis motivates a simple sequential multiplicity assignment scheme, bit-level generalized minimum
distance decoding. The proposed BGMD outperforms most of the MAS's in the literature for RS codes in a wide
range of rates over various channels both in terms of performance and complexity. Due to its simplicity, the
performance of BGMD can also be tightly bounded using order statistics based upper bounds even in high SNR's
over an AWGN channel. The proposed BGMD has potential applications in decoding RS codes in practical recording
systems and RS outer codes in concatenated systems.

\section*{Acknowledgment}
The authors are grateful to N.~Ratnakar, R.~Koetter, J.~Justesen and I.~Djurdjevic for many inspiring
discussions and insightful suggestions. We also would like to thank the associate editor Hans-Andrea Loeliger for handling the review of this paper and anonymous reviewers for their valuable comments and suggestions which significantly improve the presentation of this paper.

\appendices

\section{Derivation of the Bit-level Radius of a 1-bit Flipped BSC}
\label{pdx:1bit_bsc}

Suppose there are $e \le N$ 1-bit flipped symbols. In the MAS, we assign $M$ to the received vector and
$t M$ to the 1-bit flipped neighbors. As $M \rightarrow \infty$, the score and cost are:

\begin{align}
\label{align:score_1bit_bsc}S = (N-e)M + e M t\\
\label{align:cost_1bit_bsc}C = \frac{N}{2}[M^2(1+m t^2)(1+o(1))]
\end{align}
Plugging the score and cost into (\ref{eqn:sufficient}), we get:

\begin{align}
\label{align:sufficient_bsc_a}[(N-e)+e t] M &> \sqrt{(K-1)N(1+m t^2)} M\\
\label{align:sufficient_bsc_b}e &< \frac{N-\sqrt{N(K-1)(1+m t^2)}}{1-t}
\end{align}

For RS codes of rate $R < \frac{1}{1+m}+\frac{1}{N}$, we have $(K-1)(1+m) < N$. Setting $t = 1$ in
(\ref{align:sufficient_bsc_a}), the inequality becomes independent of $e$ and is always satisfied. In this case,
the transmitted codeword will always be on the list.

For higher rate RS codes, $t$ is optimized to maximize the right hand side (RHS) of
(\ref{align:sufficient_bsc_b}). This problem is equivalent to maximizing the slope between a given point $(1,
N)$ and a point on the hyperbola $\frac{y^2}{N(K-1)}-m x^2 = 1$, within the range $0 \le x \le 1$ and
$y~\ge~\sqrt{N(K-1)}$, which is nothing but the tangent to the hyperbola. For the tangential point $(x_0, y_0)$,
we have the following relationships:

\begin{align}
\label{eqn:relation2} \frac{y_0^2}{N(K-1)}&-mx_0^2 = 1\\
\label{eqn:relation1} \frac{dy}{dx}\mid_{x = x_0} &= N(K-1)m\frac{x_0}{y_0}\\
\label{eqn:relation1b} &= \frac{N-y_0}{1-x_0}
\end{align}

From the above three equations, we can get:
\begin{align}
\label{eqn:relation_xy}
y_0 = (K-1)(m x_0+1)
\end{align}

Plugging back to (\ref{eqn:relation2}), we get
\begin{align}
\label{eqn:function_x0}
m\left[m(K-1)-N\right]x_0^2+2m(K-1)x_0-\left[N-(K-1)\right]=0
\end{align}
Since we are only interested in $x_0 \in [0, 1]$, it is easy to verify that in all cases, the solution of
(\ref{eqn:function_x0}) will be of the following form:
\begin{equation}
\label{eqn:x_0 solution} x_0 = \frac{- m (K-1)+\sqrt{\Delta}}{m(m(K-1)-N)}
\end{equation}
where $\Delta = (m (K-1))^2+(N-K+1)(m^2(K-1)-m N)$. Note that the singular point $m(K-1)-N = 0$ can be removed
by taking the limit: $[m(K-1)-N] \rightarrow 0$. Combing (\ref{eqn:relation1}) and (\ref{eqn:relation_xy}), the
optimal error correction radius is:

\begin{equation}
\label{eqn:slope solution}
e_{max} < \frac{N}{\frac{1}{m x_0}+1}
\end{equation}
where $x_0$ is computed in (\ref{eqn:x_0 solution}). The maximum $e_{max}$ satisfying (\ref{eqn:slope solution})
is the error correction radius of ASD algorithm under the proposed asymptotically optimal MAS over 1-bit flipped BSC and
$t = x_0$ is the optimal multiplicity coefficient.

Moreover, $\sqrt{\Delta}$ can be further bounded as follows:
\begin{equation}
\label{eqn:taylor expansion}
\sqrt{\Delta} < m(K-1)[1+\frac{1}{2} \frac{(N-K+1)(m^2(K-1)-m N)}{m^2(K-1)^2}]
\end{equation}
For high rate RS codes, $(N-K+1)(m^2(K-1)-m N) \ll (m (K-1))^2$, the left hand side (LHS) and RHS of
(\ref{eqn:taylor expansion}) becomes very close and the upper bound on $\sqrt{\Delta}$ becomes tight. Plug
(\ref{eqn:taylor expansion}) into (\ref{eqn:x_0 solution}):
\begin{equation}
\label{eqn:approx_x_0}
\tilde{x}_{0} = \frac{N-(K-1)}{2m(K-1)}
\end{equation}
Plug (\ref{eqn:approx_x_0}) into (\ref{eqn:slope solution}), we finally get:
\begin{equation}
\label{eqn:approximation} \tilde{e}_{max} = \frac{N(N-K+1)}{N+(K-1)} =
\left[N-\sqrt{N(K-1)}\right]\left(1+\frac{\sqrt{N(K-1)}-(K-1)}{N+(K-1)}\right)
\end{equation}
Note that $\tilde{e}_{max} > e_{max}$ and it serves as an upper bound on the true decoding radius. However,
(\ref{eqn:approximation}) suggests that in the 1-bit flipped BSC case, the improvement of ASD over GS algorithm
is very little for high rate RS codes. A similar result was independently obtained in \cite{justesen_isit}.

\section{Derivation of the Decoding Region of ASD over a 1-bit Flipped or Erased Channel}
\label{pdx:1bit_region}

We consider the following MAS for the 1-bit flipped or erased channel: if the symbol does not contain erased
bits, assign multiplicity $M$ to the received symbol and $M t_1$ to all 1-bit flipped neighbors. In the 1-bit
erased symbols, we assign $M t_2$ to both candidates.

Suppose we have $f$ erasures and $e$ errors. For a given $f$, an optimal MAS tries to maximize $e$.

In the infinite cost case, the score and the cost are:
\begin{eqnarray}
  S &=& (N-e-f)M + e M t_1 + f M t_2\\
  C &=& \frac{1}{2}\left[(N-f)(M^2+m M^2 t_1^2) + 2f M^2 t_2^2\right](1+o(1))
\end{eqnarray}
When a received vector is certainly decodable in the infinite cost case, (\ref{eqn:sufficient}) has to be satisfied. We
have:
\begin{eqnarray}
\label{eqn:1bit_besc}  e < \frac{N-f(1-t_2)-\sqrt{(K-1)\left[(N-f)(1+m t_1^2) + 2f t_2^2\right]}}{1-t_1}
\end{eqnarray}
When $f = 0$, (\ref{eqn:1bit_besc}) reduces to (\ref{align:sufficient_bsc_b}). Here, we only consider the
non-trivial case, $f > 0$. Define
\begin{eqnarray}
    J_1 &=& N-f(1-t_2)-\sqrt{(K-1)\left[(N-f)(1+m t_1^2) + 2f t_2^2\right]}\\
    J &=& \frac{J_1}{1-t_1}
\end{eqnarray}
We first maximize $J_1$ with respect to $t_2$. Take the derivative, we get:
\begin{eqnarray}
g(t_2) = \frac{\partial{J_1}}{\partial{t_2}} = f-\frac{4 (K-1) f t_2}{2\sqrt{(K-1)\left[(N-f)(1+m t_1^2) + 2f
t_2^2\right]}}
\end{eqnarray}
Note that $g(t_2)$ is a monotonically decreasing function, with $g(0) = f > 0$. Note that $\lim_{t_2
\rightarrow \infty}{g(t_2) > 0}$ when $f > 2(K-1)$. This suggests the proposed optimal MAS will have $t_2 \rightarrow
\infty$. In this case, $S \approx f M t_2$, $C \approx f M^2 t^2_2(1+o(1))$ and $S \ge \sqrt{2(K-1)C}$ will
always be satisfied if $f > 2(K-1)$. Therefore, when $f > 2(K-1)$, $e+f = N$ errors and erasures can be
recovered for the 1-bit flipped or erased channel, which is optimal. It can also be shown that when $f = 2(K-1)
< N$, $e+f = N$ is also achievable by properly assigning multiplicities to symbols without erasure. This is not
too surprising, since the 1-bit erased symbols are guaranteed to be error free and therefore, it worth putting
more multiplicities on 1-bit erased symbols. For high rate RS codes, we have $2(K-1)
> N$. Hence, $g(t_2)$ will have a unique zero in $t_2 \in [0, \infty)$, which maximizes $J_1$. Set
$g(t_2) = 0$, we get:
\begin{eqnarray}
    t_2 &=& \sqrt{\frac{(N-f)(1 + m t_1^2)}{4(K-1)-2 f}}\\
    J_1 &=& N-f-\sqrt{\left[(K-1)-f/2\right](N-f)(1+m t^2_1)}
\end{eqnarray}

$J$ can thus be simplified as a function of $t_1$ only as:
\begin{eqnarray}
\label{eqn:J_function}
J &=& A\times \frac{B-\sqrt{1+m t_1^2}}{1-t_1}
\end{eqnarray}
where
\begin{eqnarray}
    A &=& \sqrt{\left[(K-1)-f/2\right](N-f)}\\
    B &=& \sqrt{(N-f)/\left[(K-1)-f/2\right]}
\end{eqnarray}

(\ref{eqn:J_function}) has a similar structure to (\ref{align:sufficient_bsc_b}). When $f >
2(K-1)+\frac{4(K-1)-2N}{m-1}$, let $t_1 = 1$ and $t_{2} = \sqrt{\frac{(N-f)(1 + m)}{4(K-1)-2 f}}$, the condition
(\ref{eqn:sufficient}) will always be satisfied.

For $f \le 2(K-1)+\frac{4(K-1)-2N}{m-1}$, we apply the same technique used in (\ref{align:sufficient_bsc_b})
here, i.e., to maximize the slope between the point $(1, B)$ and a point on the hyperbola $y^2 - m x^2 = 1$ will
give the optimal multiplicity coefficient $t_1$:

\begin{eqnarray}
\label{eqn:t1opt}
t_{1, opt} = \frac{-m+\sqrt{m^2+m(m-B^2)(B^2-1)}}{m(m-B^2)}
\end{eqnarray}

The optimal $J$ as a function of $f$ is:
\begin{eqnarray}
\label{eqn:Jopt}
J_{opt}(f) = (N-f)\frac{m t_{1, opt}}{m t_{1, opt}+1}
\end{eqnarray}
Eventually, the optimal decoding region is:
\begin{eqnarray}
\label{eqn:e_opt}
    e < J_{opt}(f)
\end{eqnarray}
Any received word from the 1-bit flipped or erased channel with $e$-bit errors and $f$-bit erasures satisfying (\ref{eqn:e_opt}) is certainly decodable by ASD under the proposed optimal MAS with $t_{1, opt}$ and $t_{2, opt}$ as the optimal multiplicity coefficients respectively.

\section{Proof of Theorem \ref{thm:finite_region}}
\label{pdx:finite_region}
\begin{proof} \label{prf:finite_region}
When $e+f \le N$, the cost is

\begin{equation}
\label{eqn:cost_thm8}C = (N-f) \frac{M(M+1)}{2} + f \frac{M (M+2)}{4}
\end{equation}
which does not depend on the number of errors $e$. $T(S)$, as defined in (\ref{eqn:sufficient_form2}), is a
piecewise linear function with monotonically increasing slope. Since $T(S)$ is monotonic, we first determine the
unique interval where $T(S)$ intersects $C$, i.e., $T(a(k-1)) \le C \le T((a+1)(K-1))$. Plugging
(\ref{eqn:sufficient_form2}) in  $T(a(K-1)) \le C$, we get an upper bound on $a$:

\begin{equation}
a \le \frac{-1+\sqrt{1+\frac{8C}{K-1}}}{2}
\end{equation}
with $C$ defined in (\ref{eqn:cost_thm8}). The integer solution of $a$ is:
\begin{equation}
\label{eqn:hat_a}
\hat{a} = \lfloor \frac{-1+\sqrt{1+\frac{8C}{K-1}}}{2} \rfloor
\end{equation}
The threshold of the score can then determined by

\begin{equation}
\label{eqn:threshold_s}
S^{*} = T^{-1}(C) = \frac{C}{\hat{a}+1}+\frac{\hat{a}}{2}(K-1)
\end{equation}
where $T^{-1}(C)$ is the inverse function of $T(S)$.

The received word is certainly decodable by ASD if $S > S^{*}$, where $S = (N-e-f/2)M$. Therefore, we have the final
decoding region as follows:
\begin{equation}
\label{eqn:finite_region} e < N-\frac{f}{2}-\frac{\hat{a}(\hat{a}+1)(K-1)/2+C}{M(\hat{a}+1)}
\end{equation}
where $C$ and $\hat{a}$ are defined in (\ref{eqn:cost_thm8}) and (\ref{eqn:hat_a}) respectively.
\end{proof}

\section{Computation of the Frame Error Rate Upper Bound of BGMD Decoding}
\label{pdx:bgmd_bound}

We give a detailed description of the procedure to compute an upper bound on the FER of BGMD decoder. This upper
bound is an extension of the GMD bound \cite{agrawal_gmd} for binary linear block codes with a bounded distance
decoder (BDD). Without loss of generality, we assume that the all-zero codeword is transmitted. Assuming that
BPSK is the modulation scheme and that a zero is mapped to a channel symbol $+1$, the received value for the
$i$th bit is $r_i = 1 + n_i$, where $n_i \sim {\cal N}(0,N_0/2)$.

Let $f(x,N_0) = \frac{1}{\sqrt{\pi N_0}} e^{-\frac{x^2}{N_0}}$ be the probability density function (PDF) of a
Gaussian random variable (RV) with mean zero and variance $N_0/2$. Then, the cumulative density function (CDF)
of this Gaussian RV is given by:
\begin{equation}
\label{eqn:Qx} Q(x,N_0) = \int_{x}^{\infty}{f(t,N_0)dt}
\end{equation}

The probability that one bit is in error can therefore be expressed as:
\begin{equation}
\label{eqn:Pb} P_b = Q(1,N_0).
\end{equation}

Let $f_{\alpha}^{e}$ and $f_{\alpha}^c$ be the PDF's of $|r_i|$
given that $r_i \leq 0$ and $r_i > 0$, respectively. It is shown
in \cite{agrawal_gmd} that $f_{\alpha}^{e}$ and $f_{\alpha}^c$ are
given by
\begin{align}
\label{align:err_pdf} f_{\alpha}^{e} = \frac{f(x+1)}{Q(1,N_0)}u(x)\\
\label{align:cor_pdf} f_{\alpha}^{c} =
\frac{f(x-1)}{1-Q(1,N_0)}u(x)
\end{align}
where $u(x)$ is a step function.

Therefore, the corresponding CDF's are:
\begin{align}
\label{align:err_cdf}
F_{\alpha}^{e} &= \frac{Q(1,N_0)-Q(x+1,N_0)}{Q(1,N_0)}u(x)\\
\label{align:cor_cdf} F_{\alpha}^{c} &=
\frac{1-Q(1,N_0)-Q(x-1,N_0)}{1-Q(1,N_0)}u(x)
\end{align}

Assume there are $i$ erroneous bits in the received vector. Order the received bits according to their
reliability values in decreasing order. Let $\beta_j(i)$ be the $j^{th}$ ordered reliability value in $i$
erroneous bits. That is $\beta_{1}(i) \ge \beta_{2}(i) \ge \cdots \ge \beta_{i}(i)$. On the other hand, there
are $n-i$ correct bits. Define $\gamma_{l}(n-i)$ as the $l^{th}$ value after ordering. We have $\gamma_{1}(i)
\ge \gamma_{2}(i) \ge \cdots \ge \gamma_{n-i}(n-i)$. The density of $\beta_{j}(i)$ and $\gamma_{l}(n-i)$ can be
derived using the order statistics as in \cite{papoulis}:
\begin{align}
\label{align:osd_beta} f_{\beta_{j}(i)}(x) &= \frac{i!}{(j-1)!(i-j)!}
\left[1-F_{\alpha}^{e}(x)\right]^{j-1}f_{\alpha}^{e}(x)\left[F_{\alpha}^{e}(x)\right]^{i-j}\\
\label{align:osd_gamma} f_{\gamma_{l}(n-i)}(x) &= \frac{(n-i)!}{(l-1)!(n-i-l)!}
\left[1-F_{\alpha}^{c}(x)\right]^{l-1}f_{\alpha}^{c}(x)\left[F_{\alpha}^{c}(x)\right]^{n-i-l}
\end{align}

Hence, the probability that the event $\{\beta_{j}(i) \ge
\gamma_{l}(n-i)\}$ occurs can be evaluated by the following double
integral:

\begin{equation}
\label{eqn:beta_gamma} P \left( \beta_{j}(i) \ge \gamma_{l}(n-i)
\right) = \int_{0}^{\infty}{f_{\gamma_{l}(n-i)}(x)
\int_{x}^{\infty}{f_{\beta_{j}(i)}(y) \ dy \ dx}}
\end{equation}

The performance of BGMD decoding can be bounded as follows:
\begin{equation}
\label{eqn:bgmd_ub} P_{BGMD} \le P_{ML}+P_{list} \approx P_{List}
\end{equation}

$P_{List}$ can be computed using the first order approximation in \cite{agrawal_gmd}. The basic idea is that for
a specified number of errors in the received vector, the actual FER of BGMD is upper bounded by the FER of an
error and erasure decoder with a fixed but optimized number of erasures in the LRB's. $P_{List}$ can be
expressed as:
\begin{align}
\label{align:p_list1} P_{list} &\le& \sum_{i = e_{max, M}+1}^{f_{max, M}}{P_b^{i}(1-P_b)^{n-i} \min_{(e, f) \in
D(M)}} { P((e+1) \text{errors in} \text{ the ($N-f$) MRB's}) +\sum_{i = f_{max,
M}+1}^{n}{P_b^{i}(1-P_b)^{n-i}}}\\
\label{align:p_list2} &=& \sum_{i = e_{max, M}+1}^{f_{max, M}}{P_b^{i}(1-P_b)^{n-i} \min_{(e, f) \in D(M)}} { P
\left( \beta_{e+1}(i) \ge \gamma_{n-e-f}(n-i) \right)+\sum_{i = f_{max, M}+1}^{n}{P_b^{i}(1-P_b)^{n-i}}}
\end{align}
where $D(M)$ is the set of all error and erasure pairs $(e, f)$ that is within the decoding region of the
proposed MAS for a specified multiplicity parameter $M$, as characterized in Theorem~\ref{thm:finite_region}. $f_{max, M}$
and $e_{max, M}$ are the maximum number of erasures and errors such that $(0, f_{max, M})$ and $(e_{max, M}, 0)$
still belong to $D(M)$.

\bibliographystyle{IEEEtran}
\bibliography{jiang} 

\end{document}